\documentclass[sigconf]{acmart}
\usepackage{booktabs} 
\usepackage{graphicx}  

\usepackage{amssymb}
\usepackage{bm}
\usepackage{amsmath}
\usepackage{verbatim}
\usepackage[ruled,vlined]{algorithm2e}
\usepackage{subfigure}
\usepackage{natbib}
\usepackage[Symbol]{upgreek}
\usepackage{tabularx}
\usepackage{multirow}
\usepackage{stfloats}
\usepackage{url}
\usepackage{diagbox}
\newcommand{\partitle}[1]{\vspace{2mm}\noindent \textbf{#1.}}

\setcopyright{rightsretained}

\begin{document}

\copyrightyear{2020} 
\acmYear{2020} 
\setcopyright{acmcopyright}
\acmConference[SIGIR '20]{Proceedings of the 43rd International ACM SIGIR Conference on Research and Development in Information Retrieval}{July 25--30, 2020}{Virtual Event, China}
\acmBooktitle{Proceedings of the 43rd International ACM SIGIR Conference on Research and Development in Information Retrieval (SIGIR '20), July 25--30, 2020, Virtual Event, China}
\acmPrice{15.00}
\acmDOI{10.1145/3397271.3401155}
\acmISBN{978-1-4503-8016-4/20/07}
\fancyhead{}

\title{Sampler Design for Implicit Feedback Data by Noisy-label Robust Learning}

\author{Wenhui Yu}
\thanks{School of Software, Tsinghua National Laboratory for Information Science and Technology}
\affiliation{%
  \institution{Tsinghua University}
  \city{Beijing}
  \state{China}
}
\email{yuwh16@mails.tsinghua.edu.cn}

\author{Zheng Qin}
\authornote{The corresponding author.}
\affiliation{%
  \institution{Tsinghua University}
  \city{Beijing}
  \state{China}
}
\email{qingzh@mail.tsinghua.edu.cn}


\begin{abstract}

Implicit feedback data is extensively explored in recommendation as it is easy to collect and generally applicable. However, predicting users' preference on implicit feedback data is a challenging task since we can only observe positive (voted) samples and unvoted samples. It is difficult to distinguish between the negative samples and unlabeled positive samples from the unvoted ones. Existing works, such as Bayesian Personalized Ranking (BPR), sample unvoted items as negative samples uniformly, therefore suffer from a critical noisy-label issue. To address this gap, we design an adaptive sampler based on noisy-label robust learning for implicit feedback data.

To formulate the issue, we first introduce Bayesian Point-wise Optimization (BPO) to learn a model, e.g., Matrix Factorization (MF), by maximum likelihood estimation. We predict users' preferences with the model and learn it by maximizing likelihood of observed data labels, i.e., a user prefers her positive samples and has no interests in her unvoted samples. However, in reality, a user may have interests in some of her unvoted samples, which are indeed positive samples mislabeled as negative ones. We then consider the risk of these noisy labels, and propose a Noisy-label Robust BPO (NBPO). NBPO also maximizes the observation likelihood while connects users' preference and observed labels by the likelihood of label flipping based on the Bayes' theorem. In NBPO, a user prefers her true positive samples and shows no interests in her true negative samples, hence the optimization quality is dramatically improved. Extensive experiments on two public real-world datasets show the significant improvement of our proposed optimization methods.

\end{abstract}

%
%

\ccsdesc[500]{Information systems~Collaborative filtering}
\ccsdesc[500]{Information systems~Social recommendation}
\ccsdesc[500]{Information systems~Recommender systems}
\ccsdesc[300]{Human-centered computing~Social recommendation}

\keywords{Negative Sampling, Noisy-label Robust Learning, Bayesian Point-wise Optimization, Collaborative Filtering, Item Recommendation.}

\maketitle





\section{Introduction}
Over the past decades, recommender systems have caught much attention and gained significant accuracy improvement. Recommender systems have been widely known as one of the key technologies for various online services like E-commerce and social media sites to predict users' preference based on their interaction histories. Two kinds of data are widely used to represent the interaction histories, explicit feedback data and implicit feedback data. Explicit feedback data is like numerical ``multi-class'' scores that users rate for each interacted item, and implicit feedback data is like ``purchase'' or ``browse'' in E-commerce sites, ``like'' in social media sites, ``click'' in advertisement, etc. It is the binary data indicating if a user has interacted with a certain item. In real-world applications, implicit feedback data is easier to collect and more generally applicable. However, it is more challenging to analyze, since there are only positive samples and unvoted samples while we cannot distinguish between the negative samples and unlabeled positive samples from the unvoted ones.

Bayesian Personalized Ranking (BPR) method \cite{BPR,VBPR,AES} is a widely used optimization criterion on implicit feedback data due to its outstanding performance. However there is a critical issue, all unvoted samples are regarded as negative equally in BPR. In fact, a user did not vote an item may not because he/she dislikes it, but just because he/she has not seen it yet. These positive samples are mislabeled as negative ones in existing sampling strategies, which leads to a serious noisy-label problem. To improve sampling quality, we explore the noisy-label robust learning in recommendation tasks. For each unobserved user-item interaction, we also label it as negative samples yet estimate the possibility of being mislabeled. The likelihood of observation is factorized into the likelihood of true labels and the likelihood of label noise. We learn the likelihood of true label and the likelihood of label noise jointly by maximizing the likelihood of observation. We can learn the true labels from the contaminated observed labels in this way, and make prediction with the true label we get.

However, there are several obstacles: (1) The most extensively used optimization method, BPR, is non-extendable to the noisy-label robust version. To deal with this issue, we introduce a \textbf{B}ayesian \textbf{P}oint-wise \textbf{O}ptimization (\textbf{BPO}) as our basic optimization method. (2) The space cost of saving label flipping probabilities for all unobserved user-item interactions is unacceptable. To deal with this issue, we argue that the probability matrix is a low-rank matrix thus can be maintained by Matrix Factorization (MF). (3) The log-likelihood is extensively used as maximum likelihood estimator while the log surrogate function does not work well in this situation. We design a new surrogate function and derivation strategy to address this issue.

As we mentioned above, BPR is incompatible with noisy-label robust learning due to the form of pairwise learning, therefore we introduce a point-wise optimization method BPO to learn models, which is the maximum posterior estimator derived by the Bayesian analysis. Similar with BPR, BPO aims to estimate the model parameters by maximizing the likelihood of observations hence the model tends to predict users' preference that conforms to the observed labels. Then, we take the label noise into account, and propose the \textbf{N}oisy-label robust \textbf{BPO} (\textbf{NBPO}). NBPO also maximizes the likelihood of observed sample labels while connects the true label likelihood with the observed label likelihood by the label noise likelihood. In NBPO, models are supervised by the predicted true labels thus preference prediction tends to conform to the estimated true labels. 

Finally, we validate the effectiveness of our proposed method by comparing it with several baselines on the \textit{Amazon.Electronics} and \textit{Movielens} datasets. Extensive experiments show that we improve the performance significantly by exploring noisy-label robust learning.

Specifically, our main contributions are summarized as follows: 

\begin{itemize}
{\item We propose the noisy-label robust sampler, NBPO, by taking the noisy label probabilities into consideration. We represent and train the probabilities of noisy labels with a form of MF to reduce the space and time cost.}

{\item To learn the model, we propose a novel optimization method with surrogate likelihood function and surrogate gradient. The parameters are updated by stochastic gradient descent (SGD).}

{\item We devise comprehensive experiments on two real-world datasets to demonstrate the effectiveness of our proposed methods. Codes for this paper are available on \url{https://github.com/Wenhui-Yu/NBPO}.}
\end{itemize}

\section{Related Work}
\label{sec:related_work}
After collaborative filtering (CF) model was proposed \cite{MF,item-based,user-based}, recommender systems have gained great development. Modern recommender systems uncover the underlying latent factors that encode the preference of users \cite{MF,BPR,Pairwise}. Among various CF methods, MF, \cite{MF,MF0,MF2}, a special type of latent factor models, is a basic yet the most effective recommender model. Recent years, many variants have been proposed to strengthen the presentation ability of recommendation models. \cite{Who,ChenXu,All_at_once} proposed tensor factorization models for context-aware recommendation. \cite{VBPR,Image_based,AES,text1,Key_Frame} leveraged various side information to recommend by incorporating features into basic models. \cite{O2,DCF} proposed fast algorithms to provide online recommendation. \cite{NCF,NFM,DeepFM} explored neural networks to learn user and item embeddings and how to combine them. \cite{cc,he_Attentive,he_conver,LCFN} leveraged several advanced networks, such as the attention neural network, convolutional neural network, and graph convolutional neural network, to enhance the representation ability. Though widely explored, recommendation on implicit feedbacks is still a challenging task due to poor negative sampling.

\subsection{Sampling on Implicit Feedback Data}
There is a large quantity of literature focusing on optimization on implicit feedback data \cite{BPR,WBPR,socialBPR,PU_BGD,trinityBPR,groupBPR,itemgroupBPR}. \citet{BPR} proposed BPR to optimize recommendation models on implicit feedback data, in which the item recommendation is treated as a ranking task rather than a regression task. BPR treats unvoted samples as negative samples equally and aims to maximize the likelihood of pairwise preference over positive samples and negative samples. As we argued that many unvoted samples are in fact unlabeled positive samples, BPR suffers from the noisy-label problem, i.e., some of the samples are indeed positive while labeled as ``0''. To enhance the performance on implicit feedback data, many research efforts focus on high-quality negative sampling strategies. 

\citet{PU_BGD} argued that selecting negative samples randomly in stochastic gradient descent (SGD) leads to noise in training and deteriorates the performance of the model. To address this gap, all unvoted entities are sampled as negative samples. In this way, only the noise caused by randomly sampling are handled, yet the \textit{noisy-label} problem in implicit feedback data still exists, since unvoted items are still sampled as the negative uniformly. \citet{WBPR} proposed a weighted BPR (WBPR) method which weights each negative item with a confidence coefficient. They argued that popular samples that unvoted by a certain user are more likely to be the real negative samples since they are unlikely to be neglected. Authors then devised a weight coefficient based on items' popularity and sampled all negative items non-uniformly. However, this weight strategy is empirical and impacts from users are ignored. 

To improve sampling quality, some literature designed samplers with collaborative information. It is assumed that all users are independent in BPR, \citet{groupBPR} tried to relax this constraint and proposed a method called group preference-based BPR (GBPR), which models the preference of user groups. \citet{itemgroupBPR} constructed the preference chain of item groups for each user. \citet{CPLR} uncovered the potential (unvoted) positive samples with collaborative information and enhanced BPR with the preference relationship among positive samples, potential positive samples, and negative ones. They also calculated a weight for each potential positive sample based on the similarity to the positive sample and finally proposed a collaborative pairwise learning to rank (CPLR) model. CPLR alleviates the noisy-label problem by uncovering and weighting the potential positive samples. However it is also empirical and the memory-based part of the model does not jointly work well with the model-based part in some cases. \citet{SPLR} mined the collaborative information by a high-level approach to uncover the potential positive samples, yet the eigen-decomposition of two large Laplacian matrices are computationally expensive.

There are also some efforts improving sampling quality with additional information. \citet{viewBPR} used browsing history to enrich positive samples. \citet{listrank,Listwise} constructed the list-wise preference relationship among items with the explicit feedback data. \citet{dynamicBPR,improve_pairwise} proposed dynamic negative sampling strategies to maximize the utility of a gradient step by choosing ``difficult'' negative samples. \citet{uninteresting} utilized both implicit and explicit feedback data to improve the quality of negative sampling. \citet{TDAR} replaced negative sampling by transfer learning.

In this paper, we explore the noisy-label robust regression technique to devise an adaptive sampler for negative samples. We weight samples with the probabilities of noisy labels and learn the probabilities jointly with the model. Compared with existing work, our weight mechanism is data-driven and more comprehensive: we weight all user-item tuples and the weight strategy is based on the Bayes formula rather than simple multiplication.

\subsection{Noisy-label Robust Learning}
Noisy-label issue is a critical issue in supervised machine learning tasks. Label noise misleads models and deteriorates the performance. There is an increasing amount of research literature that aims to address the issues regarding to learning from samples with noisy class label assignments \cite{label_noise_robust,PUlearning1,PUlearning2,PU_rec}.

\citet{label_noise_robust} proposed a noisy-label robust regression, which tries to learn the classifier jointly with estimating the label flipping probability. The likelihoods of real labels and of observed labels are connected by the flipping probability. Authors finally maximized the likelihood of observation to estimate all parameters. Positive and unlabeled (PU) data can be regarded as a kind of noisy-label data, in which we mainly consider the probability that positive samples are mislabeled as negative ones. \citet{PUlearning3} proposed an active learning algorithm for PU data, which works by separately estimating probability density of positive and unlabeled points and then computing expected value of informativeness to get rid of a hyperparameter and have a better measure of informativeness. \citet{PUlearning2} proposed a cost-sensitive classifier, which utilizes a non-convex loss to prevent the superfluous penalty term in the objective function. \citet{PU_rec1} proposed a matrix completion method for PU data, which can be used in recommendation tasks. However, the impact of different users and items on label flipping probabilities are neglected, different samples share the same probabilities.

In recommendation field, noisy-label problem is more serious than other supervised learning fields. That is because in real-world scenarios, users only voted a small proportion of their interested items thus most of the positive items are labeled as negative samples mistakenly. In this paper, We explore noisy-label robust learning to devise adaptive sampler for implicit feedback data, and our label flipping probabilities are sample-specific. We also propose effective learning strategy for our method.

\section{Preliminaries}
\label{sec:preliminaries}
In this section we introduce preliminaries of noisy-label robust learning and BPO. Bold uppercase letters refer to matrices. For example, ${\bm{{\rm A}}}$ is a matrix, ${\bm{{\rm A}}}_i$ is the $i$-th row of ${\bm{{\rm A}}}$, ${\bm{{\rm A}}}_{*j}$ is the $j$-th column of ${\bm{{\rm A}}}$, and ${\bm{{\rm A}}}_{ij}$ is the value at the $i$-th row and the $j$-th column of ${\bm{{\rm A}}}$. Bold lowercase letters refer to vectors. For example, ${\bm{{\rm a}}}$ is a vector, ${\bm{{\rm a}}}_i$ is the $i$-th value of the vector, and ${\bm{{\rm a}}}^{(i)}$ is the $i$-th vector. Italic letters refer to numbers. In this paper, we use $\tilde{y}^{(n)}$/$\tilde{\bm{{\rm R}}}_{ui}$ to indicate the observed label, use $y^{(n)}$/${\bm{{\rm R}}}_{ui}$ to indicate the true label and use $\hat{y}^{(n)}$/$\hat{\bm{{\rm R}}}_{ui}$ to indicate the prediction returned by the model.

\subsection{Noisy-label Robust Learning}
\label{subsec:nlrl}
Consider a training dataset $\mathcal{D}$ containing $N$ samples, $\mathcal{D}$ $=$ $\{({\bm{{\rm x}}}^{(1)},$ $\tilde{y}^{(1)}),$ $\cdots,$ $({\bm{{\rm x}}}^{(N)},\tilde{y}^{(N)})\}$, where ${\bm{{\rm x}}}^{(n)}\in \mathbb{R}^K$ is a $K$-dimensional feature and $\tilde{y}^{(n)}\in \{0,1\}$ is a binary observed label (containing noise). In the conventional logistic regression, the log likelihood is define as:
\begin{small}
\begin{eqnarray}
\label{equ:likelihood1}
\left.\begin{aligned}
\!\!\!\mathcal{L}({\bm{{\rm w}}})\!=\!\!\!\sum_{n=1}^N\! \big[\tilde{y}^{(n)}\! \log p(\tilde{y}^{(n)}\!\!=\!1|{\bm{{\rm x}}}^{(n)}\!,\!{\bm{{\rm w}}}) \!+\! (1\!-\!\tilde{y}^{(n)}) \log p(\tilde{y}^{(n)}\!\!=\!0|{\bm{{\rm x}}}^{(n)},\!{\bm{{\rm w}}})\big],
\end{aligned}
\right.
\end{eqnarray}
\end{small}where ${\bm{{\rm w}}}$ is the parameter vector of the classifier. Presuming all observed labels are correct, we use them to supervise the model training and yield the model to predict as observed labels. The probability distribution of the observed label $\tilde{y}^{(n)}$ can be represented as $p(\tilde{y}^{(n)}=1|{\bm{{\rm x}}}^{(n)},{\bm{{\rm w}}})=\sigma({\bm{{\rm w}}}^\mathsf{T} {\bm{{\rm x}}}^{(n)})$, and $p(\tilde{y}^{(n)}=0|{\bm{{\rm x}}}^{(n)},{\bm{{\rm w}}})=1-\sigma({\bm{{\rm w}}}^\mathsf{T} {\bm{{\rm x}}}^{(n)})=\sigma(-{\bm{{\rm w}}}^\mathsf{T} {\bm{{\rm x}}}^{(n)})$. Here we leverage sigmoid function $\sigma(\;)$ as the surrogate function.

However, if label noise presents in the data, noisy-label robust learning should be leveraged to alleviate the impact of noise. We use variable $y^{(n)}$ to represent the true label of the $n$-th sample, and the probability distribution of the observed label $\tilde{y}^{(n)}$ is:
\begin{small}
\begin{flalign}
p(\tilde{y}^{(n)}=k|{\bm{{\rm x}}}^{(n)},{\bm{{\rm w}}})=\sum_{j=0}^1 p(\tilde{y}=k|y=j)p(y^{(n)}=j|{\bm{{\rm x}}}^{(n)},{\bm{{\rm w}}}), \nonumber
\end{flalign}
\end{small}where $j,k\in\{0,1\}$, $p(\tilde{y}=k|y=j)$ is the probability that a label has flipped from $j$ into the observed label $k$. Instead of Equation (\ref{equ:likelihood1}), we define the likelihood function with label noise as:
\begin{small}
\begin{eqnarray}
\label{equ:likelihood2}
\left.\begin{aligned}
\mathcal{L}({\bm{{\rm w}}})=\sum_{n=1}^N \Big[&\tilde{y}^{(n)} \log \sum_{j=0}^1 p(\tilde{y}=1|y=j)p(y^{(n)}=j|{\bm{{\rm x}}}^{(n)},{\bm{{\rm w}}}) \\
+ &(1-\tilde{y}^{(n)}) \log \sum_{j=0}^1 p(\tilde{y}=0|y=j) p(y^{(n)}=j|{\bm{{\rm x}}}^{(n)},{\bm{{\rm w}}})\Big],
\end{aligned}
\right.
\end{eqnarray}
\end{small}here we use the true labels to supervise model training thus the probability distribution of a true label $y^{(n)}$ can be presented as $p(y^{(n)}=1|{\bm{{\rm x}}}^{(n)},{\bm{{\rm w}}})=\sigma({\bm{{\rm w}}}^\mathsf{T} {\bm{{\rm x}}}^{(n)})$ and $p(y^{(n)}=0|{\bm{{\rm x}}}^{(n)},{\bm{{\rm w}}})=\sigma(-{\bm{{\rm w}}}^\mathsf{T} {\bm{{\rm x}}}^{(n)})$. We learn the parameter ${\bm{{\rm w}}}$ and the noisy-label probabilities $\{p(\tilde{y}=k|y=j)\}_{j,k\in \{0,1\}}$ by maximizing the $\mathcal{L}({\bm{{\rm w}}})$ in Equation (\ref{equ:likelihood2}), and classify a new data point depending on the probability distribution of the true label $p(y^{(n)}|{\bm{{\rm x}}}^{(n)},{\bm{{\rm w}}})$.

Comparing Equations (\ref{equ:likelihood1}) and (\ref{equ:likelihood2}) we can see that aiming to minimize the possible risk on training set, both the conventional regression and noisy-label robust regression maximize the likelihood of the observed labels. In conventional regression, the observed labels are equivalent to the true labels while in noisy-label robust regression, the observed labels and the true labels are connected by the noisy-label probability based on the Bayes' theorem. In this paper, we consider that unlabeled positive samples in implicit feedback data could be mislabeled as negative samples, thus explore noisy-label robust regression in recommendation tasks to improve the sampling quality. In existing noisy-label robust learning methods \cite{label_noise_robust,PU_rec1}, for certain $j$ and $k$, all samples share the same label flipping probability $p(\tilde{y}=k|y=j)$, while in this paper we argue that the probability of label noise varies from user to user and item to item, thus we maintain a specific label flipping probability $p(\tilde{y}^{(n)}=k|y^{(n)}=j)$ for the $n$-th sample ($n=1\cdots N$).

\subsection{Bayesian Point-wise Optimization}
\label{subsec:BPO}
Since the widely-used method BPR is incompatible to the noisy-label robust learning, we first introduce a BPO learning instead of BPR as the basic optimization method. Here, we use a binary variable matrix $\tilde{\bm{{\rm R}}} \in \mathbb{R}^{M\times N}$ to represent the observed labels, where $M$ and $N$ are numbers of users and items respectively. $\tilde{\bm{{\rm R}}}_{ui}=1$ if the user $u$ has voted the item $i$, otherwise $\tilde{\bm{{\rm R}}}_{ui}=0$. We aim to predict the missing values of $\tilde{\bm{{\rm R}}}$ (samples labeled with ``0'') by reconstructing it in a low-rank form. We use $\mathcal{R}=\{(u,i)|\tilde{\bm{{\rm R}}}_{ui}=1\}$ to represent the set of observed interactions in $\tilde{\bm{{\rm R}}}$.

The Bayes formula is used in the point-wise regression to construct the maximum posterior estimator: $p({\bm{{\rm \Theta}}}|\tilde{\bm{{\rm R}}})\propto p(\tilde{\bm{{\rm R}}}|{\bm{{\rm \Theta}}})p({\bm{{\rm \Theta}}})$, where ${\bm{{\rm \Theta}}}$ represents the parameters of an arbitrary model (e.g., MF). We assume all samples are independent to each other hence the likelihood function $p(\tilde{\bm{{\rm R}}}|{\bm{{\rm \Theta}}})$ can be rewritten as a product of single probability distributions:
\begin{small}
	\begin{flalign}
	\label{equ:BPO1}
	p(\tilde{\bm{{\rm R}}}|{\bm{{\rm \Theta}}})=&\prod_{u,i} p(\tilde{\bm{{\rm R}}}_{ui}|{\bm{{\rm \Theta}}})=\prod_{(u,i)\in \mathcal{R}}p(\tilde{\bm{{\rm R}}}_{ui}=1|{\bm{{\rm \Theta}}})\prod_{(u,i)\notin \mathcal{R}}p(\tilde{\bm{{\rm R}}}_{ui}=0|{\bm{{\rm \Theta}}}) \nonumber\\
	=&\prod_{(u,i)\in \mathcal{R}}\sigma(\hat{\bm{{\rm R}}}_{ui}({\bm{{\rm \Theta}}}))\prod_{(u,i)\notin \mathcal{R}}\sigma(-\hat{\bm{{\rm R}}}_{ui}({\bm{{\rm \Theta}}}))
	\end{flalign}
\end{small}where $\hat{\bm{{\rm R}}}$ is the prediction returned by the model. Equation (\ref{equ:BPO1}) gives the formula of $p(\tilde{\bm{{\rm R}}}|{\bm{{\rm \Theta}}})$, and now we also give the formula of $p({\bm{{\rm \Theta}}})$. Assume that the random variables ${\bm{{\rm \Theta}}}$ follow normal distribution: ${\bm{{\rm \thetaup}}}\sim \mathcal{N}(0,{\bm{{\rm \Sigma}}})$, where ${\rm \bm \thetaup}$ is a column vector concatenated by all columns of ${\bm{{\rm \Theta}}}$, and ${\bm{{\rm \Sigma}}}$ is the variance-covariance matrix of ${\rm \bm \thetaup}$. the probability density of ${\bm{{\rm \Theta}}}$ is: $$p({\bm{{\rm \Theta}}})=(2\pi)^{-\frac{n}{2}}|{\bm{{\rm \Sigma}}}|^{-\frac{1}{2}}\exp\left(-\frac{{\rm \bm \thetaup}^\mathsf{T}{\bm{{\rm \Sigma}}}^{-1}{\rm \bm \thetaup}}{2}\right),$$ where $n$ is the element number of ${\bm{{\rm \Theta}}}$ (also of ${\rm \bm \thetaup}$). To reduce the number of unknown hyperparameters, we set ${\bm{{\rm \Sigma}}}=\frac{{\bm{{\rm I}}}}{\lambda_{\bm{{\rm \Theta}}}} $, and we then have $\ln p({\bm{{\rm \Theta}}}) = \frac{n}{2}(\ln\lambda_{\bm{{\rm \Theta}}}-\ln 2\pi) -\frac{\lambda_{\bm{{\rm \Theta}}}}{2} \| {\bm{{\rm \Theta}}} \|_F^2$. $| \; |$ and $\| \; \|_F$ in aforementioned formulas indicate the determinant and the Frobenius norm of the matrix, respectively. The objective function is the log likelihood of parameters given observations:
\begin{small}
	\begin{eqnarray}
	\label{equ:BPO2}
	\left.\begin{aligned}
	\mathcal{L}({\bm{{\rm \Theta}}})&=\ln p({\bm{{\rm \Theta}}}|\tilde{\bm{{\rm R}}})=\ln p(\tilde{\bm{{\rm R}}}|{\bm{{\rm \Theta}}})+\ln p({\bm{{\rm \Theta}}})\\
	&=\!\!\sum_{(u,i)\in \mathcal{R}}\!\!\ln \sigma(\hat{\bm{{\rm R}}}_{ui}({\bm{{\rm \Theta}}}))
	\!+\!\!\sum_{(u,i)\notin \mathcal{R}}\!\!\ln \sigma(-\hat{\bm{{\rm R}}}_{ui}({\bm{{\rm \Theta}}}))
	\!-\!\frac{1}{2}\lambda_{\bm{{\rm \Theta}}} \| {\bm{{\rm \Theta}}} \|_F^2 + C,
	\end{aligned}
	\right.
	\end{eqnarray}
\end{small}where $C=\frac{n}{2}(\ln\lambda_{\bm{{\rm \Theta}}}-\ln 2\pi)$ is a constant irrelevant to ${\bm{{\rm \Theta}}}$. We learn the parameter ${\bm{{\rm \Theta}}}$ by maximizing the likelihood function $\mathcal{L}({\bm{{\rm \Theta}}})$ in Equation (\ref{equ:BPO2}).

\section{Noisy-label Robust Sampler}
\label{sec:nois_label_robust_recommendation}
In this section, we propose our noisy-label robust recommendation optimization, NBPO, by incorporating the noisy-label robust learning into BPO. We first give the formulation and then the learning strategy.

\subsection{NBPO Formulation}
\label{subsec:NBPO_formulaization}
As represented in Equation (\ref{equ:likelihood2}), we need to maintain a noisy-label probability set $\{p(\tilde R=k| R=j)\}_{k,j\in\{0,1\}}$ in conventional noisy-label robust regression, which contains all transition probabilities from $j$ to $k$. Noting since the probabilities are defined for all samples, we use $\tilde R$/$R$ to indicate an arbitrary element in matrix $\tilde{\bm{{\rm R}}}$/${\bm{{\rm R}}}$. Since we consider that the implicit feedback data in recommendation context is a kind of PU data, we set $p(\tilde R=1| R=0)=0$ and $p(\tilde R=0| R=0)=1$ directly, and only consider $p(\tilde R=1| R=1)$ and $p(\tilde R=0| R=1)$. As $p(\tilde R=0| R=1)+p(\tilde R=1| R=1)=1$, we only need to learn one noisy-label probability $p(\tilde R=0| R=1)$ in NBPO. 

In conventional noisy-label robust regression introduced in Subsection \ref{subsec:nlrl}, the noisy-label probabilities are not sample-specific. However, in recommendation scenarios, the probability of an item being neglected $p(\tilde R=0| R=1)$ is user- and item-sensitive, i.e., $p(\tilde R=0| R=1)$ varies with different items and different users. For example, popular items are less likely to be missed \cite{WBPR}, and users spend more time browsing items would have a less probability to miss what they like. It also depends on user habits and item properties. Considering aforementioned reasons, we learn different noisy-label probabilities for different samples, i.e., we maintain a $M\times N$ probability matrix in NBPO. In this paper, we use ${\bm{{\rm \Gamma}}}\in \mathbb{R}^{M\times N}$ to denote the noisy-label probabilities: $\sigma({\bm{{\rm \Gamma}}}_{ui})=p(\tilde{\bm{{\rm R}}}_{ui}=0|{\bm{{\rm R}}}_{ui}=1)$. Now we rewrite the likelihood function $p(\tilde{\bm{{\rm R}}}|{\bm{{\rm \Theta}}})$ in Equation (\ref{equ:BPO1}) and log likelihood function $\mathcal{L}({\bm{{\rm \Theta}}})$ in Equation (\ref{equ:BPO2}).

The likelihood function $p(\tilde{\bm{{\rm R}}}|{\bm{{\rm \Gamma}}},\!{\bm{{\rm \Theta}}})$ is:
\begin{small}
\begin{eqnarray}
\label{equ:NBPO_likelihood}
\left.\begin{aligned}
p(\tilde{\bm{{\rm R}}}|&{\bm{{\rm \Gamma}}},\!{\bm{{\rm \Theta}}})=\!\prod_{u,i} p(\tilde{\bm{{\rm R}}}_{ui}|{\bm{{\rm \Gamma}}},\!{\bm{{\rm \Theta}}})=\!\!\!\prod_{(u,i)\in \mathcal{R}}\!\!\!p(\tilde{\bm{{\rm R}}}_{ui}\!=\!1|{\bm{{\rm \Gamma}}},\!{\bm{{\rm \Theta}}})\!\!\!\prod_{(u,i)\notin \mathcal{R}}\!\!\!p(\tilde{\bm{{\rm R}}}_{ui}\!=\!0|{\bm{{\rm \Gamma}}},\!{\bm{{\rm \Theta}}})\\
=&\prod_{(u,i)\in \mathcal{R}}p(\tilde{\bm{{\rm R}}}_{ui}=1|{\bm{{\rm R}}}_{ui}=1)p({\bm{{\rm R}}}_{ui}=1|{\bm{{\rm \Theta}}})\times \\
&\prod_{(u,i)\notin \mathcal{R}}\Big[p({\bm{{\rm R}}}_{ui}=0|{\bm{{\rm \Theta}}})+p(\tilde{\bm{{\rm R}}}_{ui}=0|{\bm{{\rm R}}}_{ui}=1)p({\bm{{\rm R}}}_{ui}=1|{\bm{{\rm \Theta}}})\Big]\\
=&\prod_{(u,i)\in \mathcal{R}}\!\!\sigma(\!-{\bm{{\rm \Gamma}}}_{ui})\sigma(\hat{\bm{{\rm R}}}_{ui}({\bm{{\rm \Theta}}}))\!\!\prod_{(u,i)\notin \mathcal{R}}\!\!\Big[\sigma(\!-\hat{\bm{{\rm R}}}_{ui}({\bm{{\rm \Theta}}}))\!+\!\sigma({\bm{{\rm \Gamma}}}_{ui})\sigma(\hat{\bm{{\rm R}}}_{ui}({\bm{{\rm \Theta}}}))\Big].
\end{aligned}
\right.
\end{eqnarray}
\end{small}Comparing Equations (\ref{equ:BPO1}) and (\ref{equ:NBPO_likelihood}), we come to the conclusion that when learning the model, we want the preference predictions $\hat{\bm{{\rm R}}}$ to be supervised by observed labels $\tilde{\bm{{\rm R}}}$ in BPO, yet by predicted true labels ${\bm{{\rm R}}}$ in NBPO. The observed labels and true labels are linked by noisy-label probabilities ${\bm{{\rm \Gamma}}}$. The log likelihood function $\mathcal{L}({\bm{{\rm \Gamma}}},{\bm{{\rm \Theta}}})$ is:
\begin{small}
	\begin{eqnarray}
	\label{equ:NBPO_loglikelihood}
	\left.
	\begin{aligned}
		&\mathcal{L}({\bm{{\rm \Gamma}}},{\bm{{\rm \Theta}}})\!=\!\ln p(\tilde{\bm{{\rm R}}}|{\bm{{\rm \Gamma}}},\!{\bm{{\rm \Theta}}})\!+\!\ln p({\bm{{\rm \Gamma}}})\!+\!\ln p({\bm{{\rm \Theta}}})=\!\!\!\sum_{(u,i)\in \mathcal{R}}\!\!\! \ln \sigma(-{\bm{{\rm \Gamma}}}_{ui}) \sigma(\hat{\bm{{\rm R}}}_{ui}({\bm{{\rm \Theta}}})) \\
		&+\!\!\!\sum_{(u,i)\notin \mathcal{R}}\!\!\!\ln \Big[ \sigma(-\hat{\bm{{\rm R}}}_{ui}({\bm{{\rm \Theta}}})) \!+\! \sigma({\bm{{\rm \Gamma}}}_{ui}) \sigma(\hat{\bm{{\rm R}}}_{ui}({\bm{{\rm \Theta}}})) \Big] \!-\! \frac{1}{2}\lambda_{\bm{{\rm \Gamma}}} \| {\bm{{\rm \Gamma}}} \|_F^2 \!-\!\frac{1}{2}\lambda_{\bm{{\rm \Theta}}} \| {\bm{{\rm \Theta}}} \|_F^2.
	\end{aligned}
	\right.
	\end{eqnarray}
\end{small}Expectation-Maximization algorithm (EM) is widely used to solve noisy-label robust learning tasks. Nevertheless EM is inefficient to solve large-scale latent variables and is not extensible to deep models, thus we aim to design a SGD-based method. However, experiments show that optimizing Equation (\ref{equ:NBPO_loglikelihood}) with SGD is rather suboptimal (shown in Figure \ref{fig:RQ3}). Inspired by EM, we construct the lower bound of Equation (\ref{equ:NBPO_loglikelihood}):
\begin{small}
	\begin{eqnarray}
	\label{equ:NBPO_lowerBound}
	\left.
	\begin{aligned}
	&\mathcal{L}({\bm{{\rm \Gamma}}},{\bm{{\rm \Theta}}})\!\geq\!\mathcal{L}_{LB}({\bm{{\rm \Gamma}}},{\bm{{\rm \Theta}}})\!=\sum_{(u,i)\in \mathcal{R}}\!\!\! \ln \sigma(-{\bm{{\rm \Gamma}}}_{ui}) \sigma(\hat{\bm{{\rm R}}}_{ui}({\bm{{\rm \Theta}}}))\!\\ &+\!\!\!\!\!\sum_{(u,i)\notin \mathcal{R}}\!\!\! \Big[\! \ln\sigma(-\hat{\bm{{\rm R}}}_{ui}({\bm{{\rm \Theta}}})) \!+\! \ln\sigma({\bm{{\rm \Gamma}}}_{ui}) \sigma(\hat{\bm{{\rm R}}}_{ui}({\bm{{\rm \Theta}}})) \Big] \!-\! \frac{1}{2}\lambda_{\bm{{\rm \Gamma}}} \| {\bm{{\rm \Gamma}}} \|_F^2 \!-\!\frac{1}{2}\lambda_{\bm{{\rm \Theta}}} \| {\bm{{\rm \Theta}}} \|_F^2.
	\end{aligned}
	\right.
	\end{eqnarray}
\end{small}In Equation (\ref{equ:NBPO_lowerBound}), we use the Jensen inequality\footnote{If $\lambda_1,\cdots, \lambda_n$ are positive numbers which sum to 1 and $f(\;)$ is a real continuous function that is concave, we have the Jensen inequality: $f(\sum_i \lambda_i x_i)\geq \sum_i \lambda_i f(x_i)$. In Equation (\ref{equ:NBPO_lowerBound}), we set $x_1 = 2\sigma(-\hat{\bm{{\rm R}}}_{ui}({\bm{{\rm \Theta}}}))$ and $x_2 = 2\sigma({\bm{{\rm \Gamma}}}_{ui}) \sigma(\hat{\bm{{\rm R}}}_{ui}({\bm{{\rm \Theta}}}))$; $\lambda_1=\lambda_2=1/2$; $f(\;) = \ln(\;)$ then we have:

\begin{small}
	\begin{flalign}
	&\ln \Big[ \sigma(-\hat{\bm{{\rm R}}}_{ui}({\bm{{\rm \Theta}}})) + \sigma({\bm{{\rm \Gamma}}}_{ui}) \sigma(\hat{\bm{{\rm R}}}_{ui}({\bm{{\rm \Theta}}})) \Big] \nonumber\\ 
	\geq &\frac{1}{2}\Big[ \ln\sigma(-\hat{\bm{{\rm R}}}_{ui}({\bm{{\rm \Theta}}})) + \ln\sigma({\bm{{\rm \Gamma}}}_{ui}) \sigma(\hat{\bm{{\rm R}}}_{ui}({\bm{{\rm \Theta}}})) +2\ln2 \Big] \nonumber\\
	\geq &\Big[\ln\sigma(-\hat{\bm{{\rm R}}}_{ui}({\bm{{\rm \Theta}}})) + \ln\sigma({\bm{{\rm \Gamma}}}_{ui}) \sigma(\hat{\bm{{\rm R}}}_{ui}({\bm{{\rm \Theta}}}))\Big]. \nonumber
	\end{flalign}
\end{small}Notice that $\ln2\geq \ln\sigma(-\hat{\bm{{\rm R}}}_{ui}({\bm{{\rm \Theta}}}))$ and $\ln2\geq \ln\sigma({\bm{{\rm \Gamma}}}_{ui}) \sigma(\hat{\bm{{\rm R}}}_{ui}({\bm{{\rm \Theta}}}))$ since $0\leq \sigma(-\hat{\bm{{\rm R}}}_{ui}({\bm{{\rm \Theta}}}))\leq 1$ and $0\leq \sigma({\bm{{\rm \Gamma}}}_{ui}) \sigma(\hat{\bm{{\rm R}}}_{ui}({\bm{{\rm \Theta}}}))\leq 1$.} to get the low bound of the likelihood function, denoted as $\mathcal{L}_{LB}({\bm{{\rm \Gamma}}},{\bm{{\rm \Theta}}})$. We optimize our model with $\mathcal{L}_{LB}({\bm{{\rm \Gamma}}},{\bm{{\rm \Theta}}})$ instead of $\mathcal{L}({\bm{{\rm \Gamma}}},{\bm{{\rm \Theta}}})$ to achieve better performance.

Obviously, NBPO suffers from an oversized-parameter issue. The $M \times N$ matrix ${\bm{{\rm \Gamma}}}$ is too large to store and to learn. To address this issue, we need to reduce the number of probability hyperparameters. We argue that since similar users/items have similar habits/properties, noisy-label probabilities of similar users/items are linearly dependent, thus ${\bm{{\rm \Gamma}}}$ is a low-rank matrix and we can represent it by a Collaborative Filtering (CF) model: $\hat{\bm{{\rm \Gamma}}}({\bm{{\rm \Phi}}})$, where $\hat{\bm{{\rm \Gamma}}}$ is the reconstruction of ${\bm{{\rm \Gamma}}}$ and ${\bm{{\rm \Phi}}}$ indicates parameters of the CF model. In NBPO, ${\bm{{\rm \Theta}}}$ is called model parameters and ${\bm{{\rm \Phi}}}$ is called optimization parameters. We use $\hat{\bm{{\rm \Gamma}}}({\bm{{\rm \Phi}}})$ to replace ${\bm{{\rm \Gamma}}}$ in Equation (\ref{equ:NBPO_lowerBound}) and $\mathcal{L}_{LB}({\bm{{\rm \Gamma}}},{\bm{{\rm \Theta}}}) = \mathcal{L}_{LB}({\bm{{\rm \Phi}}},{\bm{{\rm \Theta}}})$. We maximize Equation (\ref{equ:NBPO_lowerBound}) by mini-batch stochastic gradient descent (MSGD): ${\bm{{\rm \Phi}}} = {\bm{{\rm \Phi}}} + \eta\frac{\partial \mathcal{L}_{LB}({\bm{{\rm \Phi}}},{\bm{{\rm \Theta}}})}{\partial {\bm{{\rm \Phi}}}}$ and ${\bm{{\rm \Theta}}} = {\bm{{\rm \Theta}}} + \eta\frac{\partial \mathcal{L}_{LB}({\bm{{\rm \Phi}}},{\bm{{\rm \Theta}}})}{\partial {\bm{{\rm \Theta}}}}$.

\subsection{NBPO Learning}
\label{subsec:NBPO_learning}
To learn the model with NBPO, we still face a critical issue: when optimized by maximum likelihood estimator, log surrogate function does not fit the situation of noisy-label robust recommendation. To be specific, when learning ${\bm{{\rm \Phi}}}$ and ${\bm{{\rm \Theta}}}$ by maximizing $\mathcal{L}_{LB}({\bm{{\rm \Phi}}},{\bm{{\rm \Theta}}})$, which is represented in Equation (\ref{equ:NBPO_lowerBound}), we get:
\begin{small}
	\begin{flalign}
	&{\bm{{\rm \Phi}}}\!=\!\!\mathop{\arg\max}\limits_{\bm{{\rm \Phi}}}\!\!\sum\limits_{(\!u,i\!)\in \mathcal{R}}\!\!\ln\sigma(-\hat{\bm{{\rm \Gamma}}}_{ui}({\bm{{\rm \Phi}}}))\!+\!\!\sum\limits_{(\!u,i\!)\notin \mathcal{R}}\!\!\ln\sigma(\hat{\bm{{\rm \Gamma}}}_{ui}({\bm{{\rm \Phi}}}))\! -\! \frac{1}{2}\lambda_{\bm{{\rm \Phi}}} \| {\bm{{\rm \Phi}}} \|_F^2, \nonumber \\
	&{\bm{{\rm \Theta}}}\!=\!\mathop{\arg\max}\limits_{\bm{{\rm \Theta}}}\!\!\!\sum\limits_{(\!u,i\!)\in \mathcal{R}}\!\!\ln\sigma(\hat{\bm{{\rm R}}}_{ui}({\bm{{\rm \Theta}}}))+\!\!\!\sum\limits_{(\!u,i\!)\notin \mathcal{R}}\!\!\ln\sigma(\hat{\bm{{\rm R}}}_{ui}({\bm{{\rm \Theta}}}))(-\hat{\bm{{\rm R}}}_{ui}({\bm{{\rm \Theta}}}))\! -\! \frac{1}{2}\lambda_{\bm{{\rm \Theta}}} \| {\bm{{\rm \Theta}}} \|_F^2. \nonumber
	\end{flalign}
\end{small}It is obvious that ${\bm{{\rm \Phi}}}$ and ${\bm{{\rm \Theta}}}$ are learnt separately. $\ln(\;)$ is widely used to surrogate the likelihood function due to the special properties, for example, it converts multiplication to addition. However, this property degrades the performance of noisy-label robust learning. We want ${\bm{{\rm \Phi}}}$ to control the magnitude of gradient when update ${\bm{{\rm \Theta}}}$, while impacted by the log function, learning ${\bm{{\rm \Phi}}}$ and ${\bm{{\rm \Theta}}}$ are two independent procedures. To deal with this issue, we remove $\ln(\;)$ from $\mathcal{L}_{LB}({\bm{{\rm \Phi}}},{\bm{{\rm \Theta}}})$ in Equation (\ref{equ:NBPO_lowerBound}) to get a new surrogate likelihood function $\mathcal{L}_s({\bm{{\rm \Phi}}},\!{\bm{{\rm \Theta}}})$:
\begin{small}
\begin{eqnarray}
\label{equ:NBPO_surrogate_log_likelihood}
\left.\begin{aligned}
\mathcal{L}_s({\bm{{\rm \Phi}}},\!{\bm{{\rm \Theta}}})\!&=\!\!\!\sum_{(u,i)\in \mathcal{R}}\!\! \Big[\sigma(-\hat{\bm{{\rm \Gamma}}}_{ui}({\bm{{\rm \Phi}}})) \sigma(\hat{\bm{{\rm R}}}_{ui}({\bm{{\rm \Theta}}}))\Big]+\!\!\!\sum_{(u,i)\notin \mathcal{R}}\!\!\!\ \Big[ \sigma(-\hat{\bm{{\rm R}}}_{ui}({\bm{{\rm \Theta}}}))\\
&+\! \sigma(\hat{\bm{{\rm \Gamma}}}_{ui}({\bm{{\rm \Phi}}})) \sigma(\hat{\bm{{\rm R}}}_{ui}({\bm{{\rm \Theta}}})) \Big] \!-\! \frac{1}{2}\lambda_{\bm{{\rm \Phi}}} \| {\bm{{\rm \Phi}}} \|_F^2 \!-\!\frac{1}{2}\lambda_{\bm{{\rm \Theta}}} \| {\bm{{\rm \Theta}}} \|_F^2.
\end{aligned}
\right.
\end{eqnarray}
\end{small}

However, a new issue appears: without $\ln(\;)$, surrogate sigmoid function faces vanishing gradient problem. Different from the vanishing gradient problem in deep learning, here we use ``vanishing gradient'' to indicate the phenomenon that the gradient becomes 0 at two ends of the domain. For example, if we want to maximize $\sigma(x)$ with SGD, we update $x$ by $x=x+\eta\sigma'(x)$. However, the gradient $\sigma'(x)=\sigma(x)\sigma(-x)$ becomes 0 when $x\to -\infty$. That is, $\sigma(x)$ cannot be trained to the maximum with SGD when current $x$ is very small. To deal with this issue, we use surrogate differential operator $\frac{\partial\ln(\cdot\;)}{\partial x}$ to replace $\frac{\partial \cdot}{\partial x}$ for all sigmoid function terms in Equation (\ref{equ:NBPO_surrogate_log_likelihood}). The surrogate gradient of Equation (\ref{equ:NBPO_surrogate_log_likelihood}) is finally given by\footnote{We use $\nabla_x$ to denote the surrogate gradient with respect to $x$. The surrogate gradient of $\sigma(x)$ is $\nabla_x\sigma(x)=\frac{\partial\ln\sigma(x)}{\partial x}=\sigma(-x)$.}:
\begin{small}
\begin{eqnarray}
\label{equ:NBPO_gradient1}
\left.\begin{aligned}
\nabla_{\bm{{\rm \Phi}}}& \mathcal{L}_s({\bm{{\rm \Phi}}},\!{\bm{{\rm \Theta}}})\!=\!\!\!\sum_{(u,i)\in \mathcal{R}}\!\!\! -\sigma(\hat{\bm{{\rm \Gamma}}}_{ui}({\bm{{\rm \Phi}}})) \sigma(\hat{\bm{{\rm R}}}_{ui}({\bm{{\rm \Theta}}}))\frac{\partial \hat{\bm{{\rm \Gamma}}}_{ui}({\bm{{\rm \Phi}}})}{\partial {\bm{{\rm \Phi}}}}\\
&+\sum_{(u,i)\notin \mathcal{R}} \sigma(-\hat{\bm{{\rm \Gamma}}}_{ui}({\bm{{\rm \Phi}}})) \sigma(\hat{\bm{{\rm R}}}_{ui}({\bm{{\rm \Theta}}}))\frac{\partial \hat{\bm{{\rm \Gamma}}}_{ui}({\bm{{\rm \Phi}}})}{\partial {\bm{{\rm \Phi}}}}- \lambda_{\bm{{\rm \Phi}}} {\bm{{\rm \Phi}}}, \\
\nabla_{\bm{{\rm \Theta}}}& \mathcal{L}_s({\bm{{\rm \Phi}}},\!{\bm{{\rm \Theta}}})\!=\!\!\!\sum_{(u,i)\in \mathcal{R}}\!\!\! \sigma(-\hat{\bm{{\rm \Gamma}}}_{ui}({\bm{{\rm \Phi}}})) \sigma(-\hat{\bm{{\rm R}}}_{ui}({\bm{{\rm \Theta}}}))\frac{\partial {\bm{{\rm R}}}_{ui}({\bm{{\rm \Theta}}})}{\partial {\bm{{\rm \Theta}}}}\\
&+\!\!\!\sum_{(u,i)\notin \mathcal{R}}\!\! \Big[\!-\!\sigma(\hat{\bm{{\rm R}}}_{ui}({\bm{{\rm \Theta}}}))\!+\! \sigma(\hat{\bm{{\rm \Gamma}}}_{ui}({\bm{{\rm \Phi}}})) \sigma(\!-\hat{\bm{{\rm R}}}_{ui}({\bm{{\rm \Theta}}}))\Big]\!\frac{\partial {\bm{{\rm R}}}_{ui}({\bm{{\rm \Theta}}})}{\partial {\bm{{\rm \Theta}}}} \!-\! \lambda_{\bm{{\rm \Theta}}} {\bm{{\rm \Theta}}}.
\end{aligned}
\right.
\end{eqnarray}
\end{small}We then update ${\bm{{\rm \Phi}}}$ and ${\bm{{\rm \Theta}}}$ by ${\bm{{\rm \Phi}}} = {\bm{{\rm \Phi}}} + \eta\nabla_{\bm{{\rm \Phi}}} \mathcal{L}_s({\bm{{\rm \Phi}}},\!{\bm{{\rm \Theta}}})$ and ${\bm{{\rm \Theta}}} = {\bm{{\rm \Theta}}} + \eta\nabla_{\bm{{\rm \Theta}}} \mathcal{L}_s({\bm{{\rm \Phi}}},\!{\bm{{\rm \Theta}}})$.

\subsection{Learning MF with NBPO}

\begin{figure}[ht]
    \setlength{\abovecaptionskip}{2mm}
    \centering
    \includegraphics[scale = 0.32]{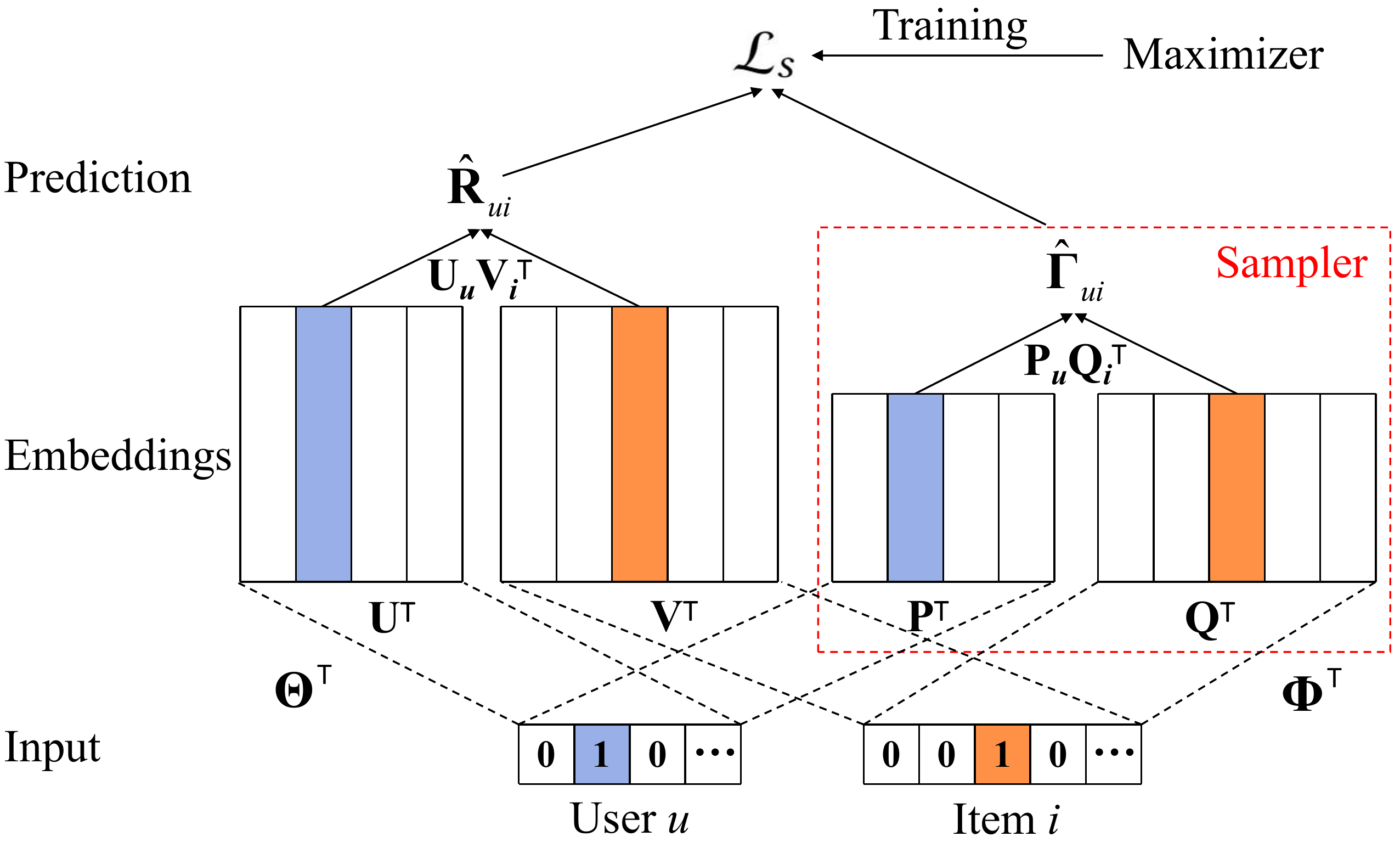}
    \caption{Illustration of our MF\_NBPO model.}
    \label{fig:illustration}
\end{figure}

To show how NBPO optimization works, we now optimize a specific model, MF, with it (denoted as MF-NBPO). MF is a basic yet very effective model. It predicts users' preference by reconstructing the purchase record matrix $\tilde{\bm{{\rm R}}}$ in a low-rank form: $\hat{\bm{{\rm R}}} = {\bm{{\rm U}}}{\bm{{\rm V}}}^\mathsf{T}$, where $\hat{\bm{{\rm R}}}$ is the reconstruction of $\tilde{\bm{{\rm R}}}$, ${\bm{{\rm U}}}\in \mathbb{R}^{M\times K}$ and ${\bm{{\rm V}}}\in \mathbb{R}^{N\times K}$ are latent factor matrices of users and items, and ${\bm{{\rm \Theta}}} = \{{\bm{{\rm U}}}, {\bm{{\rm V}}}\}$. ${\bm{{\rm U}}}_u$ indicates the latent factors which encode the preference of user $u$ and ${\bm{{\rm V}}}_i$ indicates the latent factors which encode the properties of item $i$. ${\bm{{\rm \Gamma}}}$ is also represented by MF, $\hat{\bm{{\rm \Gamma}}}={\bm{{\rm P}}}{\bm{{\rm Q}}}^\mathsf{T}$, where $\hat{\bm{{\rm \Gamma}}}$ is the reconstruction of ${\bm{{\rm \Gamma}}}$; ${\bm{{\rm P}}}\in \mathbb{R}^{M\times L}$ and ${\bm{{\rm Q}}}\in \mathbb{R}^{N\times L}$; and ${\bm{{\rm \Phi}}} = \{{\bm{{\rm P}}}, {\bm{{\rm Q}}}\}$. Figure \ref{fig:illustration} shows the structure of our MF\_NBPO model. ${\rm \bm \Theta}$ predicts the true labels and ${\rm \bm \Phi}$ predicts the label flipping probability. We then predict the observed labels with $\hat{\bm{{\rm R}}}$ and $\hat{\bm{{\rm \Gamma}}}$, and finally maximize the likelihood function shown in Equation (\ref{equ:NBPO_surrogate_log_likelihood}).

The first order partial differential in Equation (\ref{equ:NBPO_gradient1}) is:
\begin{equation}
\label{equ:NBPO_gradient2}
\!\!\!\!\begin{array}{lcl}
{\frac{\partial \hat{\bm{{\rm R}}}_{ui}({\rm \bm \thetaup})}{\partial {\rm \bm \thetaup}} \!=\! \left\{\!\!\!
\begin{array}{lcl}
{\bm{{\rm V}}}_i,\,\,{\rm if} &\!\!\!\!\!{\rm \bm \thetaup} = {\bm{{\rm U}}}_{u}\\
{\bm{{\rm U}}}_u,\,\,{\rm if} &\!\!\!\!\!{\rm \bm \thetaup} = {\bm{{\rm V}}}_{i}
\end{array}  
\right.}\!\!, \!&\!
{\frac{\partial \hat{\bm{{\rm \Gamma}}}_{ui}({\rm \bm \phiup})}{\partial {\rm \bm \phiup}} \!=\! \left\{\!\!\!
\begin{array}{lcl}
{\bm{{\rm Q}}}_i,\,\,{\rm if} &\!\!\!\!\!{\rm \bm \phiup} = {\bm{{\rm P}}}_{u}\\
{\bm{{\rm P}}}_u,\,\,{\rm if} &\!\!\!\!\!{\rm \bm \phiup} = {\bm{{\rm Q}}}_{i}
\end{array}  
\right.}\!\!.
\end{array} 
\end{equation}
Please note that $\frac{\partial \hat{\bm{{\rm R}}}_{ui}({\rm \bm \thetaup})}{\partial {\rm \bm \thetaup}}$ and $\frac{\partial \hat{\bm{{\rm \Gamma}}}_{ui}({\rm \bm \phiup})}{\partial {\rm \bm \phiup}}$ in Equation (\ref{equ:NBPO_gradient2}) refer to a certain row of $\frac{\partial \hat{\bm{{\rm R}}}_{ui}({\rm \bm \Theta})}{\partial {\rm \bm \Theta}}$ and $\frac{\partial \hat{\bm{{\rm \Gamma}}}_{ui}({\rm \bm \Phi})}{\partial {\rm \bm \Phi}}$ in Equation (\ref{equ:NBPO_gradient1}) respectively. Taking $\frac{\partial \hat{\bm{{\rm R}}}_{ui}({\rm \bm \thetaup})}{\partial {\rm \bm \thetaup}}$ as an example, it is the $u$-th row of $\frac{\partial \hat{\bm{{\rm R}}}_{ui}({\rm \bm \Theta})}{\partial {\rm \bm \Theta}}$ when ${\rm \bm \thetaup}$ is the $u$-th row of ${\rm \bm \Theta}$. 

Inspired by pairwise optimization strategy \cite{BPR,NCF}, we adapt similar negative sampling strategy in our point-wise optimization (BPO and NBPO): we do not select all unvoted samples as the negative when updating the model; we just select $\rho$ unvoted samples randomly for each positive sample, where $\rho$ is the sampling rate. With the gradient given in Equations (\ref{equ:NBPO_gradient1}) and (\ref{equ:NBPO_gradient2}), we can learn MF\_NBPO with MSGD.

To make a prediction, we use the true labels ${\bm{{\rm R}}}$ to score all items for a certain user. As aforementioned, the prediction of the true label likelihood is $\sigma(\hat{\bm{{\rm R}}})$. For a certain user $u$, we rank all items by $\sigma(\hat{\bm{{\rm R}}}_u)$ and return the top-$k$ items. Considering that $\sigma(\;)$ increases monotonically, we rank items by descending $\hat{\bm{{\rm R}}}_u={\bm{{\rm U}}}_u{\bm{{\rm V}}}^\mathsf{T}$. As we can see, parameters ${\bm{{\rm \Phi}}}$ only assistant to learn ${\bm{{\rm \Theta}}}$ while do not contribute to prediction directly.

An important merit of our NBPO is that it is updated by gradient descent method hence is scalable to deep models, which are widely used in real-world applications. In NBPO, we can choose more powerful models such as Neural Matrix Factorization (NMF) \cite{NCF} and Attentive Collaborative Filtering (ACF) \cite{he_Attentive} to construct $\hat{\bm{{\rm R}}}({\bm{{\rm \Theta}}})$ and $\hat{\bm{{\rm \Gamma}}}({\bm{{\rm \Phi}}})$. In this paper, we choose the most simple model, MF, to emphasize the effectiveness of our noisy-label robust optimization method. Learning deep models by NBPO is left to explore in the future work.

\section{Experiments}
\label{sec:experimets}
In this section, we conduct experiments to demonstrate the effectiveness of our proposed model. We report the performances of several state-of-the-art models and our model on two real-world public datasets to illustrate the precision enhancement. We focus on answering following research questions:

\vspace{2mm}
\noindent \textbf{RQ1:} How is the performance of our noisy-label robust recommendation model with point-wise optimization (NBPO)?
\vspace{1mm}

\noindent \textbf{RQ2:} How is the performance enhancement by taking noisy labels into consideration?
\vspace{1mm}

\noindent \textbf{RQ3:} How is the effectiveness of our surrogate likelihood function and surrogate gradient?

\subsection{Experimental Setup}
In this subsection, we introduce the datasets, baselines, evaluation protocols, and parameter tuning strategies.

\subsubsection{Datasets} In this paper, we adopt two real-world datasets, \textit{Amazon}\footnote{\url{http://snap.stanford.edu/data/amazon/productGraph/categoryFiles/reviews_Electronics_5.json.gz}} and \textit{Movielens}\footnote{\url{http://grouplens.org/datasets/movielens/1m/}}, to learn all models.

\begin{itemize}
    \item \textbf{Amazon.} This \textit{Amazon} dataset \cite{Amazon1,Amazon2} is the user reviews collected from the E-commerce website \textit{Amazon.com}. In this paper we adopt the \textit{Electronic} category, which contains the purchase records of electronic products in \textit{Amazon.com}. We choose the 5-core version (remove users and items with less than 5 purchase records).
    \item \textbf{MovieLens.} This \textit{Movielens} dataset \cite{Movielens} is collected through the movie website \textit{movielens.umn.edu}. This movie rating dataset has been widely used to evaluate CF models. 1M version is adapted in our experiments.
\end{itemize}

\begin{table}[ht]
    \caption{Statistics of datasets}
    \centering
    \label{tab:datasets}
    \scalebox{1}{
    \begin{tabular}{ccccc}
        \toprule[1.2pt]
                Dataset & Purchase & User & Item & Sparsity\\
        \hline
        \textit{Amazon} & 1,689,188 & 192,403 & 63,001 & 99.9861\% \\
        \textit{Movielens} & 1,000,209 & 6,040 & 3,900 & 95.7535\% \\
        \bottomrule[1.2pt]
    \end{tabular}}
\end{table}

These two datasets are all explicit feedbacks (rating data), we set the interaction $(u,i)$ as ``1'' if $u$ rated $i$ and ``0'' otherwise to construct implicit feedbacks. Table \ref{tab:datasets} shows some statistics of datasets. As shown in the table, though filtered with 5-core, the sparsity of \textit{Amazon} dataset is still extremely high, thus we filter it further with 14-core (we select 14 to balance the sparsity and the size). We split each dataset into three subsets randomly: training set (80\%), validation set (10\%), and test set (10\%). We train models on training sets, and determine all hyperparameters on validation sets, and finally report the performances on test sets. Cold items and users (items and users with no record in training set) in validation and test sets are removed.

\subsubsection{Baselines}
\label{subsec:baselines}
We adopt the following methods as baselines for performance comparison to demonstrate the feasibility and effectiveness of our model.
\begin{itemize}
\item{\textbf{ItemPop:} This method ranks items based on their popularity. It is a non-personalized method to benchmark the recommendation performances.}

\item{\textbf{ItemKNN:} This is the standard item-based CF method \cite{item-based}. We use this memory-based CF model to benchmark the performances of model-based CF models.}

\item{\textbf{BPR:} This \textbf{B}ayesian \textbf{P}ersonalized \textbf{R}anking method is the most widely used ranking-based method for implicit feedback \cite{BPR}. It regards unvoted samples as negative samples uniformly and maximizes the likelihood of users' preference over a pair of positive and negative samples.}

\item{\textbf{WBPR:} This \textbf{W}eighted \textbf{B}ayesian \textbf{P}ersonalized \textbf{R}anking me-thod \citep{WBPR} is an extension of BPR. WBPR improves the quality of negative sampling depending on the item popularity. Considering that popular items are unlikely to be neglected, WBPR gives larger confidence weights to negative samples with higher popularity.}

\item{\textbf{ShiftMC}:} \citet{PUlearning2} proposed a density function to deal with the noisy label problem. Following \cite{PUlearning2}, \citet{PU_rec1} proposed the \textbf{Shifted} \textbf{M}atrix \textbf{C}ompletion method by exploring the density function in CF model. ShiftMC is the state-of-the-art PU data-faced recommendation model.
\end{itemize}

Strictly speaking, methods proposed in our paper (BPO, NBPO) and several baselines (BPR, WBPR, ShiftMC) are optimization methods which can be used to optimize any recommendation models, such as MF, Factorization Machine (FM), Neural Collaborative Filtering (NCF), etc. In this paper, we validate the effectiveness of all optimization methods by training MF model with them, thus should be denoted as ``MF-BPR'', ``MF-WBPR'', ``MF-ShiftMC'', ``MF-BPO'', and ``MF-NBPO''. In the rest of this paper, we omit ``MF-'' for concise representation.

\begin{figure*}[ht]
    \setlength{\abovecaptionskip}{2mm}
    \centering
    \subfigure[\textit{Amazon} --- $F_1$score@$k$]{
        \includegraphics[scale = 0.225]{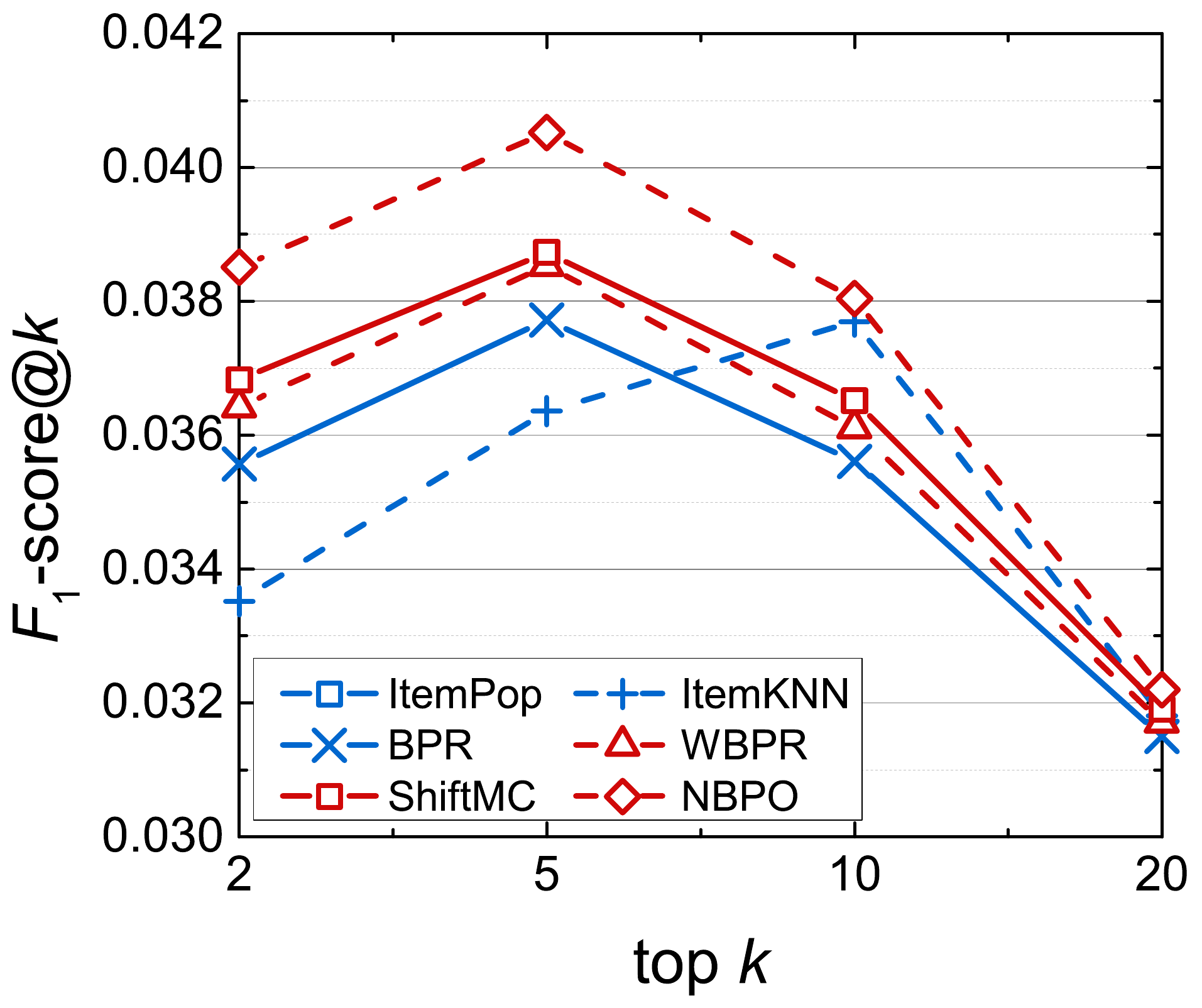}
        \label{subfig:amazon_testset_f1}
    }
    \hspace{-2mm}
    \subfigure[\textit{Amazon} --- NDCG@$k$]{
        \includegraphics[scale = 0.225]{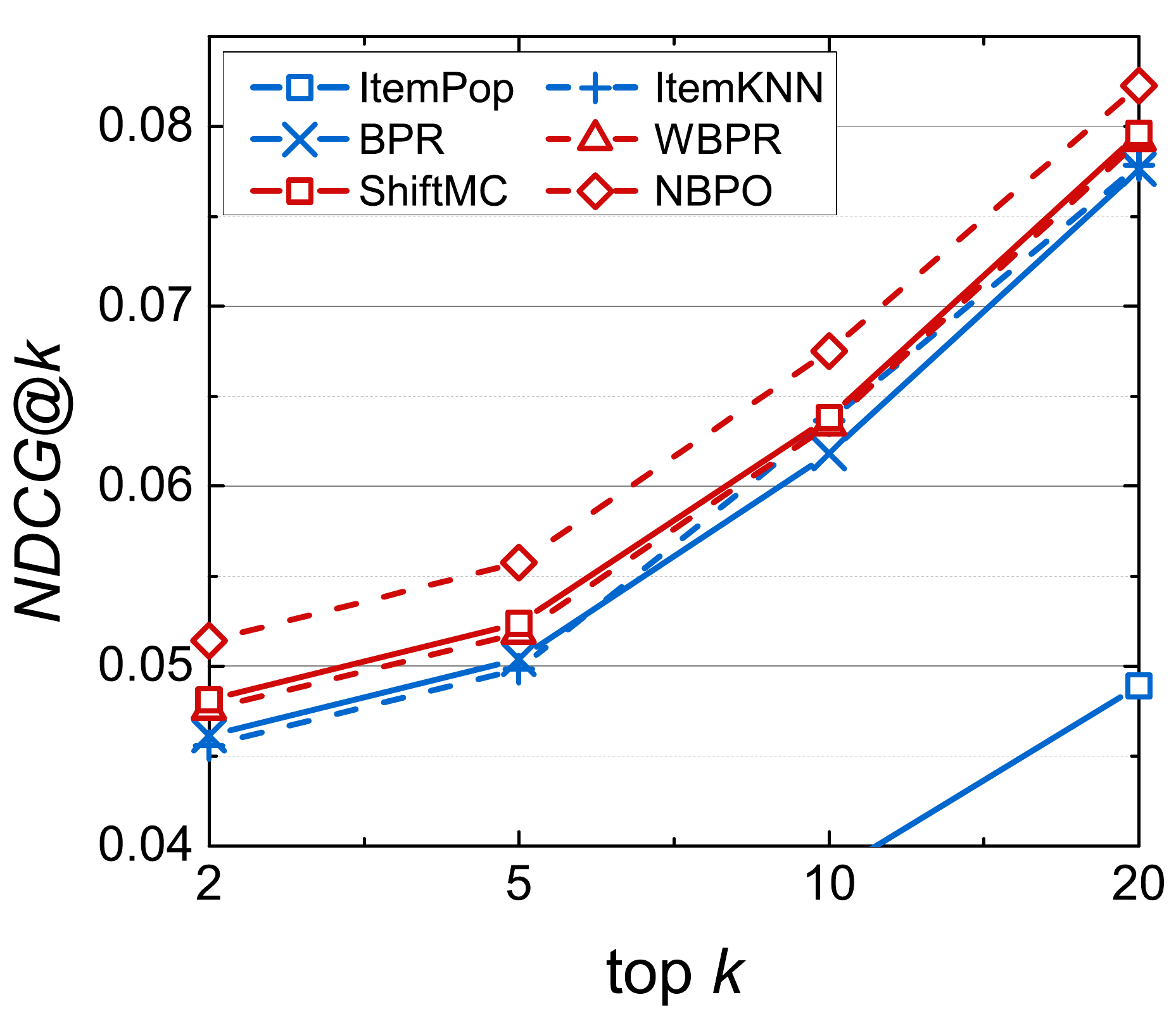}
        \label{subfig:amazon_testset_ndcg}
    }
    \hspace{-2mm}
    \subfigure[\textit{Movielens} --- $F_1$score@$k$]{
        \includegraphics[scale = 0.225]{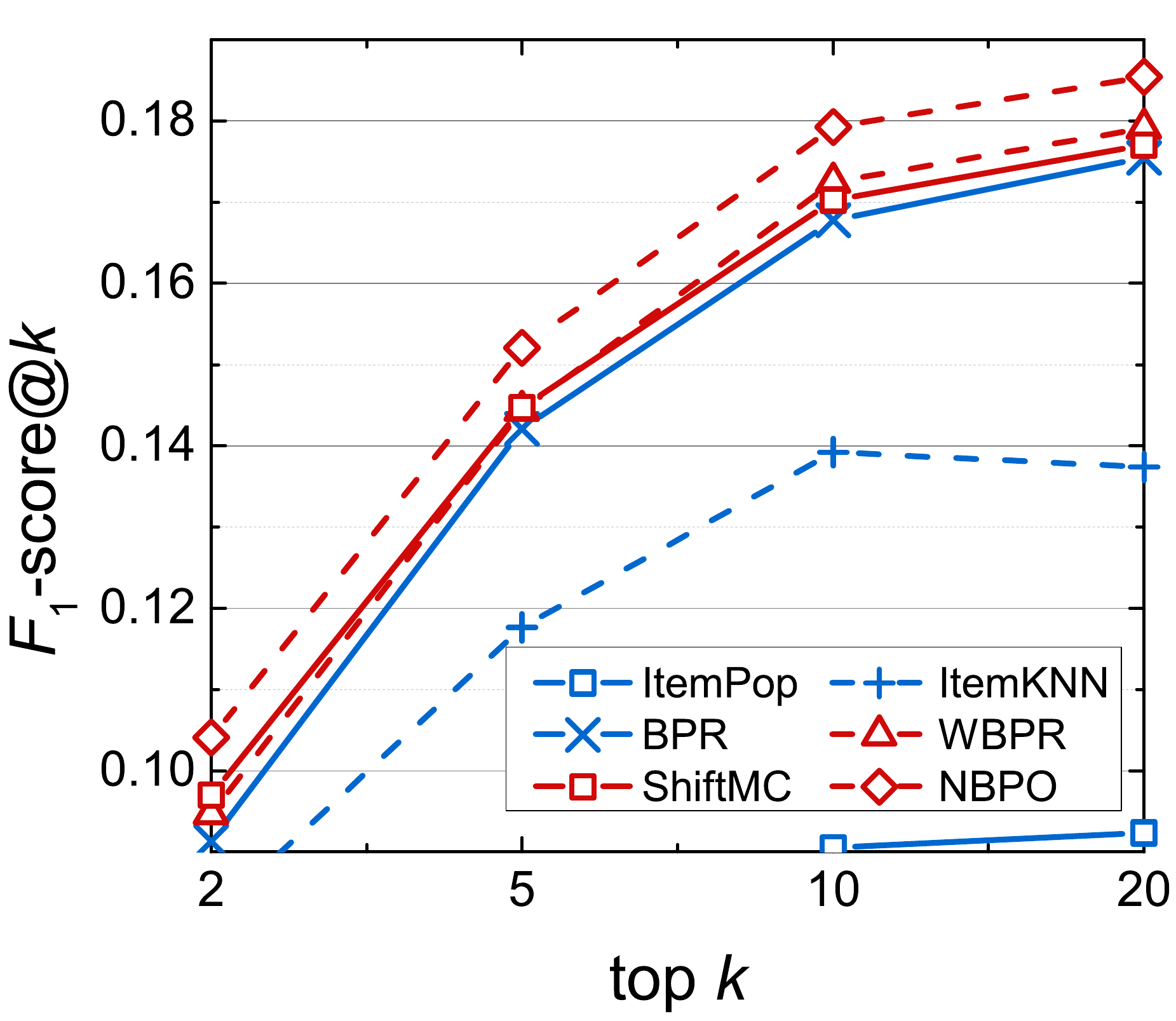}
        \label{subfig:movielens_testset_f1}
    }
    \hspace{-2mm}
    \subfigure[\textit{Movielens} --- NDCG@$k$]{
        \includegraphics[scale = 0.225]{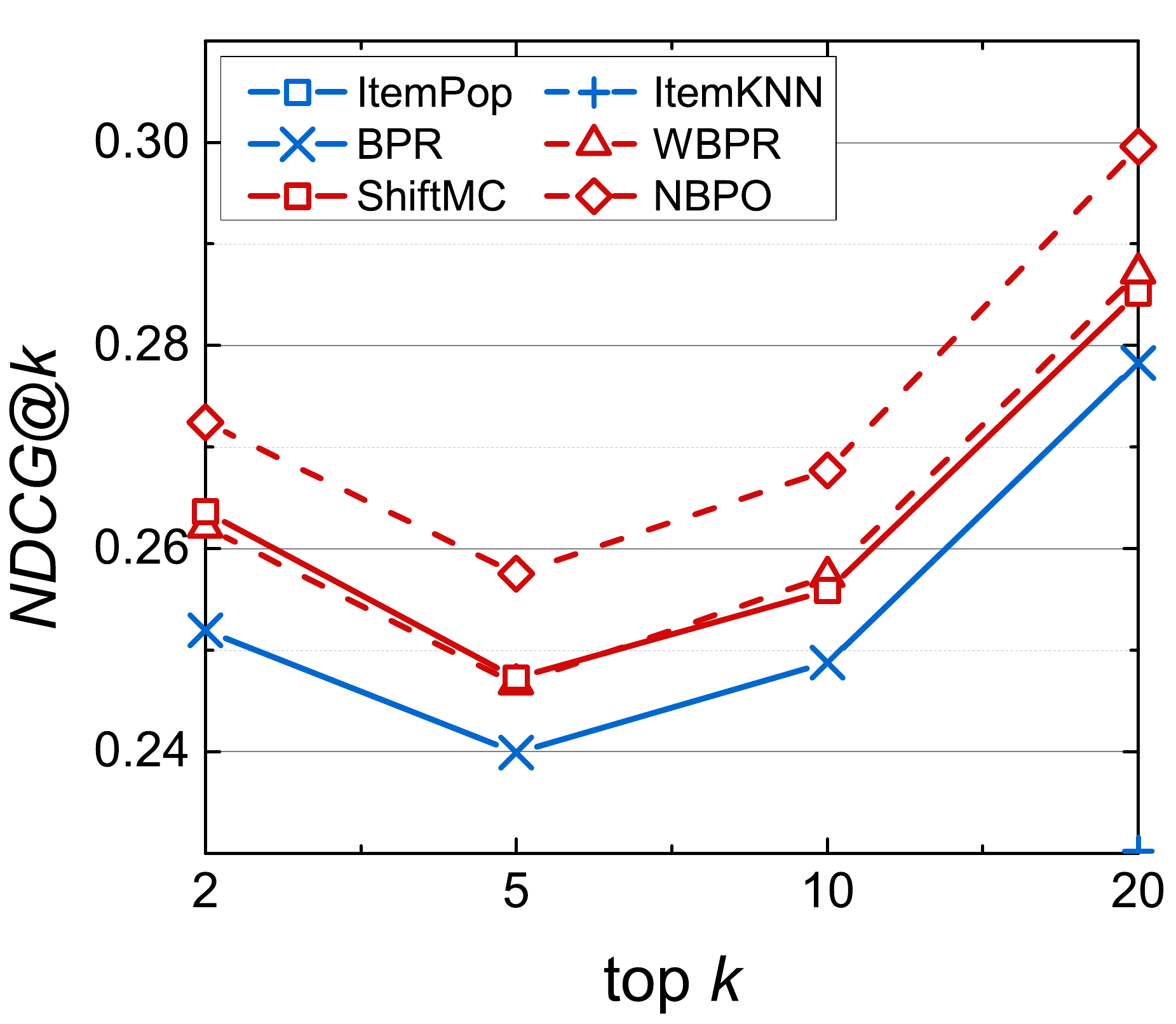}
        \label{subfig:movielens_testset_ndcg}
    }
    \caption{Recommendation performances for RQ1 (test set).}
    \label{fig:testset}
\end{figure*}

\subsubsection{Evaluation Protocols}
To evaluate the performances of our proposed model and baselines in implicit feedback context, we rank all items for each user in validation/test set and recommend the top-$k$ items to the user. We then adopt two metrics, $F_1$-score and normalized discounted cumulative gain (NDCG) to evaluate the recommendation quality. $F_1$-score, which is defined as harmonic mean of precision and recall, is extensively used to test the accuracy of a binary classifier. NDCG is a position-sensitive metric widely used to measure the ranking quality.
We recommend top-$k$ and calculate metrics for each user, and finally use the average metrics of all users to remark the performance of the models.

\subsubsection{Parameter Setting}
\label{subsec:parameter_setting}
In this subsection, we introduce the detailed parameter tuning strategy. The maximum iteration number is set to 200. In each iteration, we enumerate all positive samples and select $\rho$ negative samples for each positive one randomly to train the model and then test it. We tune all models according to the performance of recommending top-2 items in the validation set. For fair comparison, all models in our experiments are tuned with the same strategy: the learning rate $\eta$ and regularization coefficient $\lambda_{\bm{{\rm \Theta}}}$ ($\lambda_{\bm{{\rm \Phi}}}$) are determined by grid search in the coarse grain range of $\{0.001, 0.01, 0.1\} \otimes \{0.01, 0.1, 1\}$ and then in the fine grain range, which is based on the result of coarse tuning. For example, if a certain model achieves the best performance when $\eta=0.01$ and $\lambda_{\bm{{\rm \Theta}}}=0.1$, we then tune it in the range of $\{0.002, 0.005, 0.01, 0.02, 0.05\} \otimes \{0.02, 0.05, 0.1, 0.2, 0.5\}$. We set $\lambda_{\bm{{\rm \Theta}}}=\lambda_{\bm{{\rm \Phi}}}$ in NBPO in this stage, and then determine them in fine grain grid search. The batch size is determined in the range of $\{1000,2000,\cdots, 5000\}$ and the sampling rate $\rho$ is searched in the range of $\{1,2,\cdots,7\}$. We evaluate different number of latent factors $K$ and $L$ in the range of $\{10, 20, 50, 100, 200\}$ and $\{0, 1, 2, 5, 10, \cdots, 500\}$, respectively. 
We conduct our experiments by predicting top-$\{2, 5, 10, 20\}$ items to users.

\subsection{Performance of NBPO (RQ1)}
In Figure \ref{fig:testset}, we repeat each model 10 times and report average performances in our experiments. To focus on our model, curves of some uncompetitive baselines, such as ItemPop and ItemKNN, are not completely shown, or even not shown. Comparing Figures \ref{fig:testset}(a)(b) and (c)(d), it is obvious that the dataset with higher sparsity shows lower performances. ItemPop is a very weak baseline since it is very rough and simple. We can see that it cannot be shown (completely) in Figure \ref{fig:testset} due to the poor performance. Utilizing collaborative information, all personalized methods outperform ItemPop dramatically. Among these CF models, ItemKNN is a rule-based recommendation strategy thus is empirical, and it only explores one-order connections in the user-item graph. Compared with ItemKNN, model-based CF models, i.e., BPR, WBPR, ShiftMC, NBPO explore high-order collaborative information, thus gain further enhancement in most cases. An interesting observation is that in Figure \ref{subfig:amazon_testset_f1}, all learning models peak at $k=5$ while ItemKNN peaks at $k=10$. The reason may be that learning models are tuned according to $F_1$-score@2, thus may not achieve the best performance when $k$ is large. It also leads to another phenomenon: the gaps among these learning models reduces with the increasing of $k$, since they are not well-tuned for top-20 item recommendation. To get the best performance for large $k$, we can retune models according to $F_1$-score@20.

By finding credible negative samples, WBPR gains better performance than BPR. However the weight mechanism of WBPR is empirical and rough, thus the enhancement is very limited: WBPR outperforms BPR 3.75\% for the best case on $F_1$-score and 4.05\% on NDCG. Taking the label noise of the PU data into consideration, ShitfMC performs the best in baselines: it outperforms BPR by 6.26\% on $F_1$-score and 4.66\% on NDCG for the best case. However, the label noise probability is not sample-specific thus there is still room for improvement. Benefiting from the sample-specific label noise probabilities and the novel optimization strategy, the improvement of NBPO is significant: it outperforms ShitfMC 7.28\% and 6.79\% on $F_1$-score and NDCG respectively for the best case.


To make a fair comparison, we tune all models with the same strategy (please see Subsection \ref{subsec:parameter_setting}). To report the result of model tuning, we show the variation of $F_1$-score@2 with respect to different hyperparameters.

\begin{figure}[ht]
    \setlength{\abovecaptionskip}{2mm}
    \centering
    \subfigure[\textit{Amazon} --- $F_1$score@2]{
        \includegraphics[scale = 0.255]{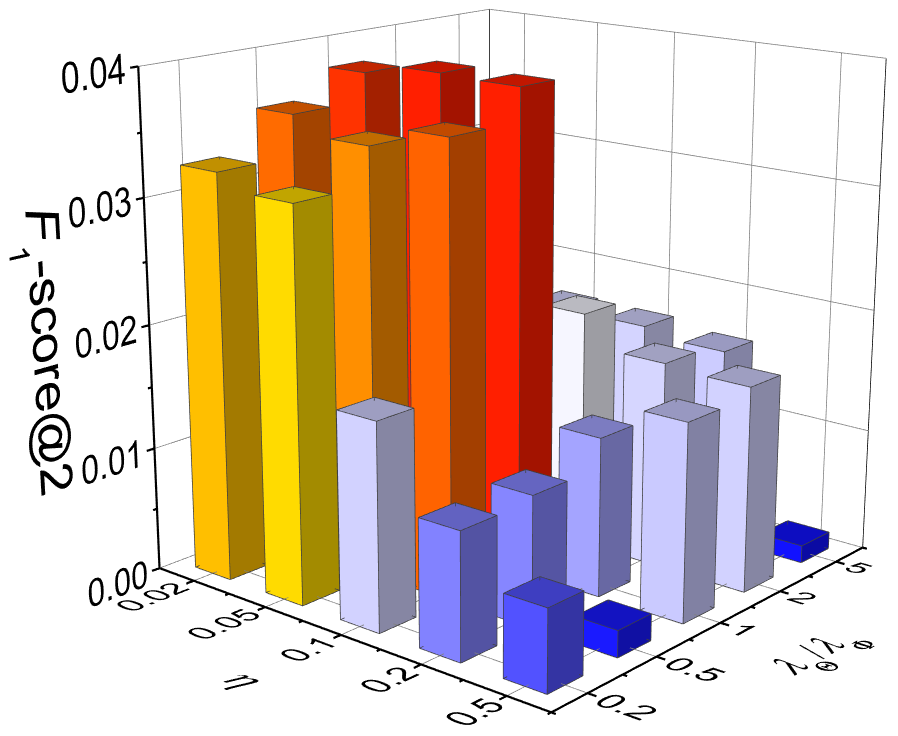}
        \label{subfig:amazon_lr_f1}
    }
    \hspace{-2mm}
    \subfigure[\textit{Movielens} --- $F_1$score@2]{
        \includegraphics[scale = 0.255]{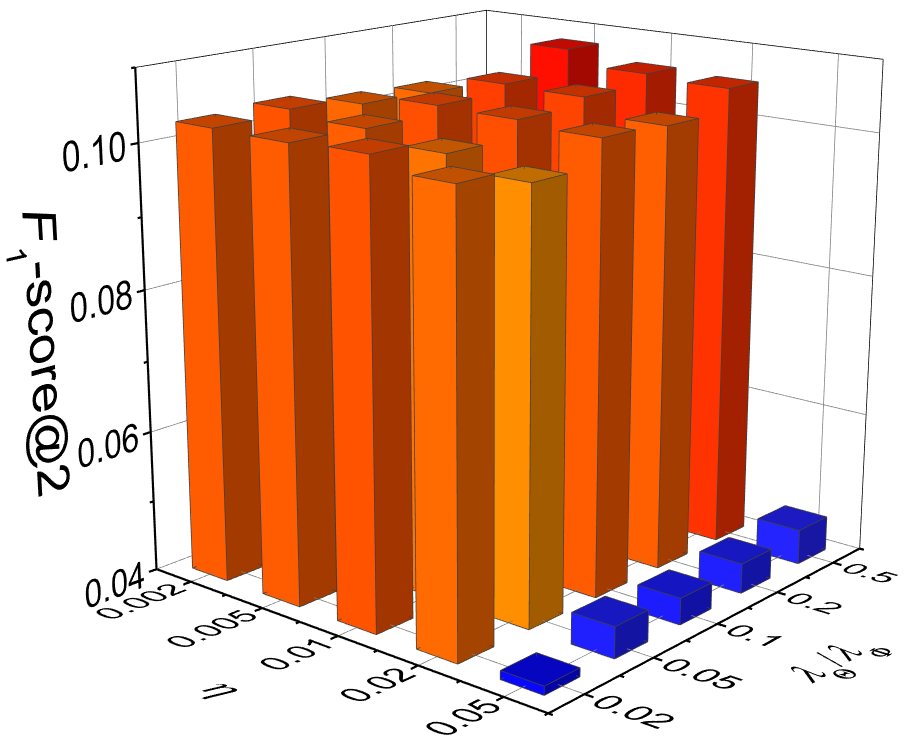}
        \label{subfig:movielens_lr_f1}
    }
    \caption{Impact of learning rate $\eta$ and regulation coefficient $\lambda_{\bm{{\rm \Theta}}}$ ($\lambda_{\bm{{\rm \Phi}}}$) (validation set).}
    \label{fig:lr}
\end{figure}

The sensitivity analysis of learning rate $\eta$ and regulation coefficient $\lambda_{\bm{{\rm \Theta}}}$ ($\lambda_{\bm{{\rm \Phi}}}$) is shown in Figure \ref{fig:lr}. To save space, we only report the fine tuning of NBPO. From Figure \ref{fig:lr} we can observe that NBPO achieves the best performance at $\eta=0.05$, $\lambda_{\bm{{\rm \Theta}}}=\lambda_{\bm{{\rm \Phi}}}=1$ on \textit{Amazon} dataset and $\eta=0.005$, $\lambda_{\bm{{\rm \Theta}}}=\lambda_{\bm{{\rm \Phi}}}=0.5$ on \textit{Movielens} dataset. We also report the best learning rate and regulation coefficient for other models: on \textit{Amazon}, BPR, WBPR, ShiftMC all achieve the best performance when $\eta=0.05$ and $\lambda_{\bm{{\rm \Theta}}}=1$, and on \textit{Movielens}, BPR, WBPR, ShiftMC achieve the best performance when $\eta=0.005$ and $\lambda_{\bm{{\rm \Theta}}}=0.2$, 0.5, 0.5, respectively.

\begin{figure}[ht]
    \setlength{\abovecaptionskip}{2mm}
    \centering
    \subfigure[\textit{Amazon} --- $F_1$score@2]{
        \includegraphics[scale = 0.21]{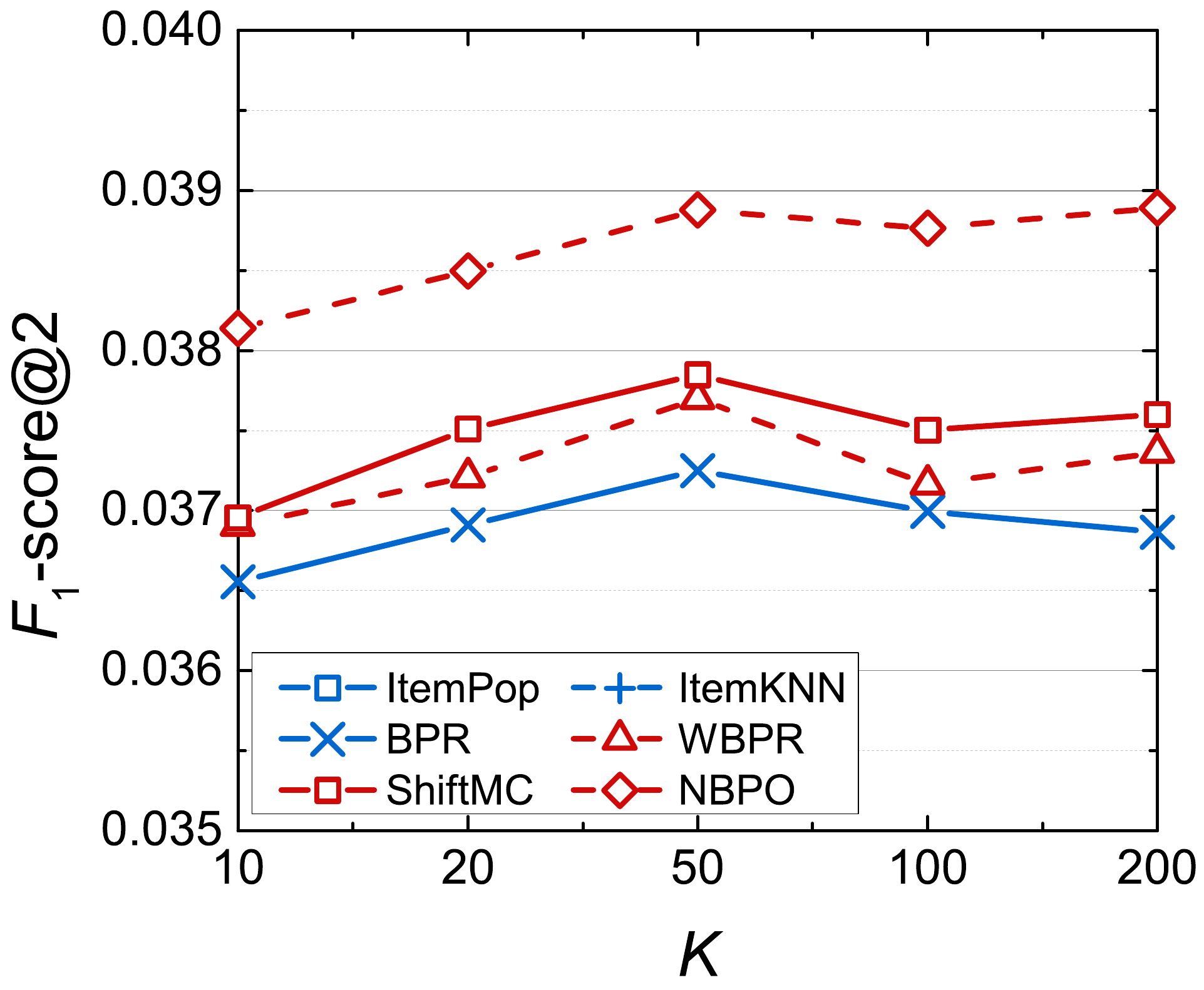}
        \label{subfig:amazon_K_f1}
    }
    \hspace{-3mm}
    \subfigure[\textit{Movielens} --- $F_1$score@2]{
        \includegraphics[scale = 0.21]{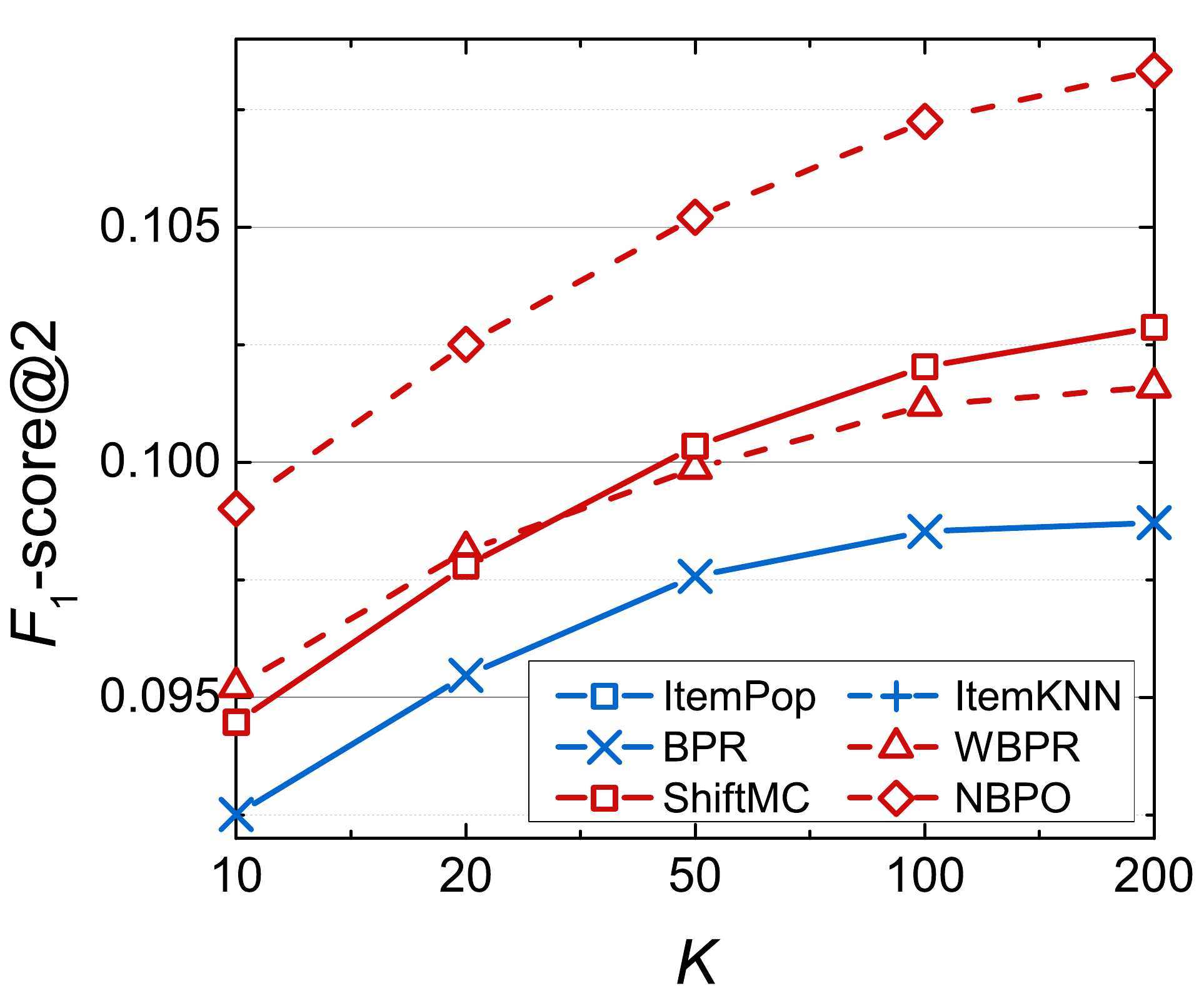}
        \label{subfig:Movielens_K_f1}
    }
    \caption{Impact of latent dimension number $K$ (validation set).}
    \label{fig:K}
\end{figure}

Models' representation abilities depend on the number of latent dimensions $K$ (In NBPO, we model users' preference with ${\bm{{\rm \Thetaup}}}$ only, and ${\bm{{\rm \Phi}}}$ is just used to help learning the model, hence the representation ability depends on $K$ rather than $L$). The impact of $K$ is represented in Figure \ref{fig:K}. Comparing Figures \ref{subfig:amazon_K_f1} and \ref{subfig:Movielens_K_f1} we can see that performances of models increase with the increasing of $K$ obviously on \textit{Movielens} dataset while keep stable on \textit{Amazon} dataset. This may be because \textit{Amazon} suffers a more serious sparsity problem, thus models easily face the overfitting problems, and stronger representation ability (larger $L$) may worsen this issue. All models achieve the best performance when $K=50$ on \textit{Amazon} dataset and $K=200$ on \textit{Movielens} dataset.

\begin{figure}[ht]
    \setlength{\abovecaptionskip}{2mm}
    \centering
    \subfigure[\textit{Amazon} --- $F_1$score@2]{
        \includegraphics[scale = 0.21]{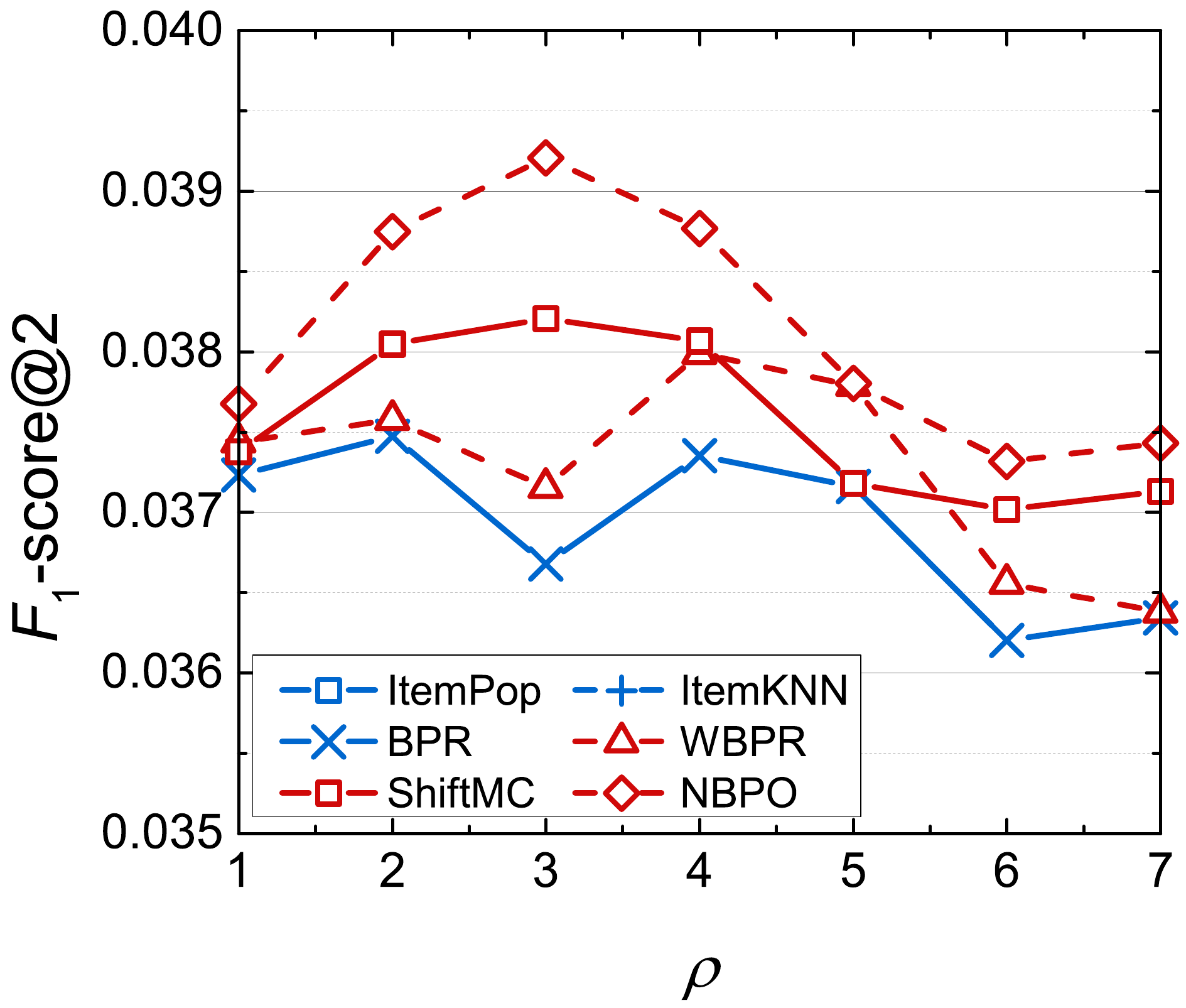}
        \label{subfig:amazon_rho_f1}
    }
    \hspace{-3mm}
    \subfigure[\textit{Movielens} --- $F_1$score@2]{
        \includegraphics[scale = 0.21]{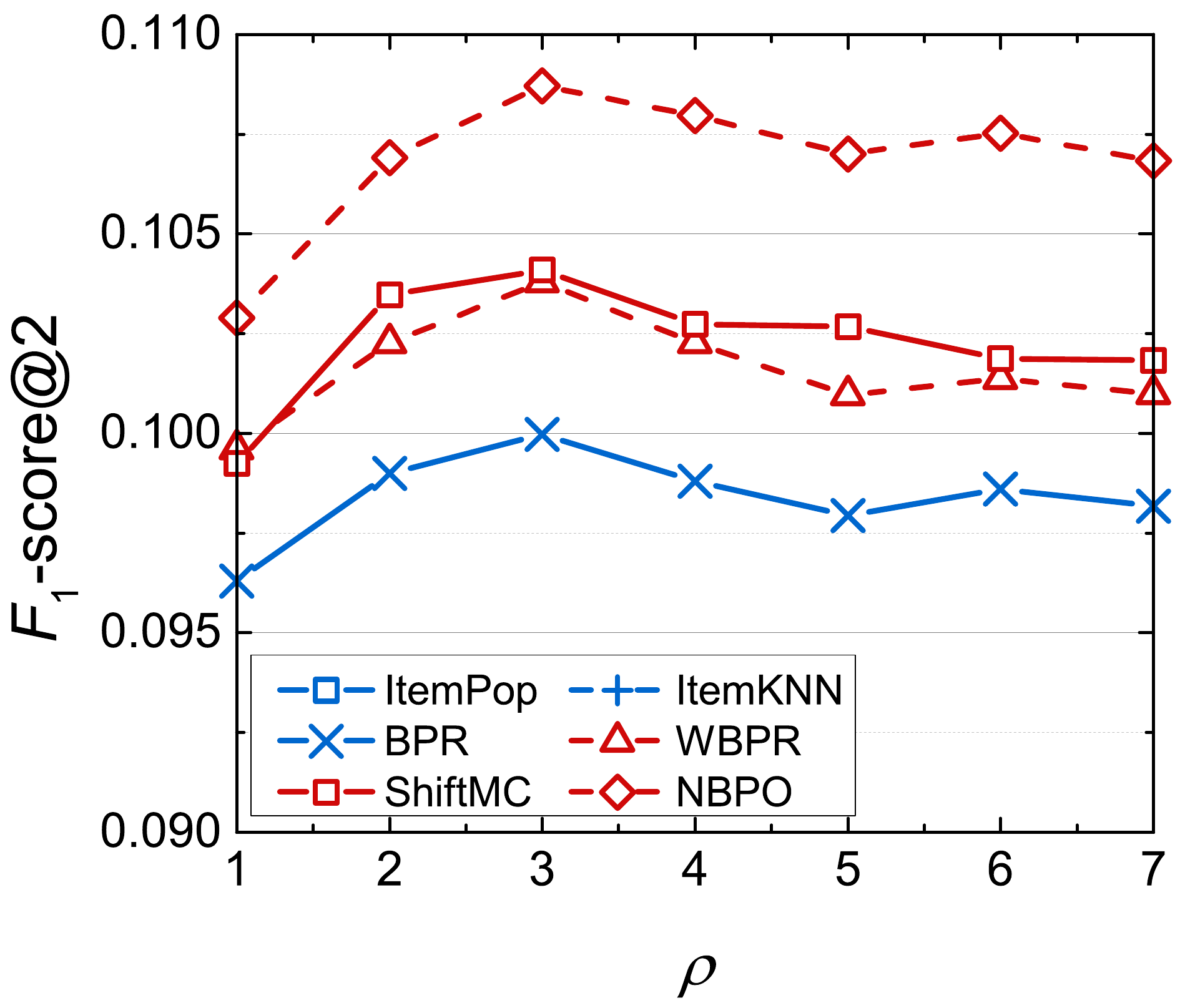}
        \label{subfig:Movielens_rho_f1}
    }
    \caption{Impact of sampling rate $\rho$ (validation set).\vspace{-1mm}}
    \label{fig:rho}
\end{figure}

We also tune $\rho$ for all models. As shown in Figure \ref{fig:rho}, BPR, WBPR, ShiftMC, NBPO perform the best when $\rho=2$, 4, 3, 3, respectively on \textit{Amazon} dataset and when $\rho=3$ on \textit{Movielens} dataset. An observation that attracts our interests is that compared with baselines, NBPO gains more improvement by tuning with $\rho$. The reason may be that the negative sampling quality is better in our NBPO model, thus sampling more negative items can boost the performance. While in BPR, sampling more negative items leads to more serious noisy-label problem, thus the improvement is limited.

\subsection{Effectiveness of Noisy-label Robust Learning in Recommendation (RQ2)}
\label{subsec:RQ2}

\begin{figure}[ht]
    \setlength{\abovecaptionskip}{2mm}
    \centering
    \subfigure[\textit{Amazon} --- $F_1$score@$k$]{
        \includegraphics[scale = 0.2]{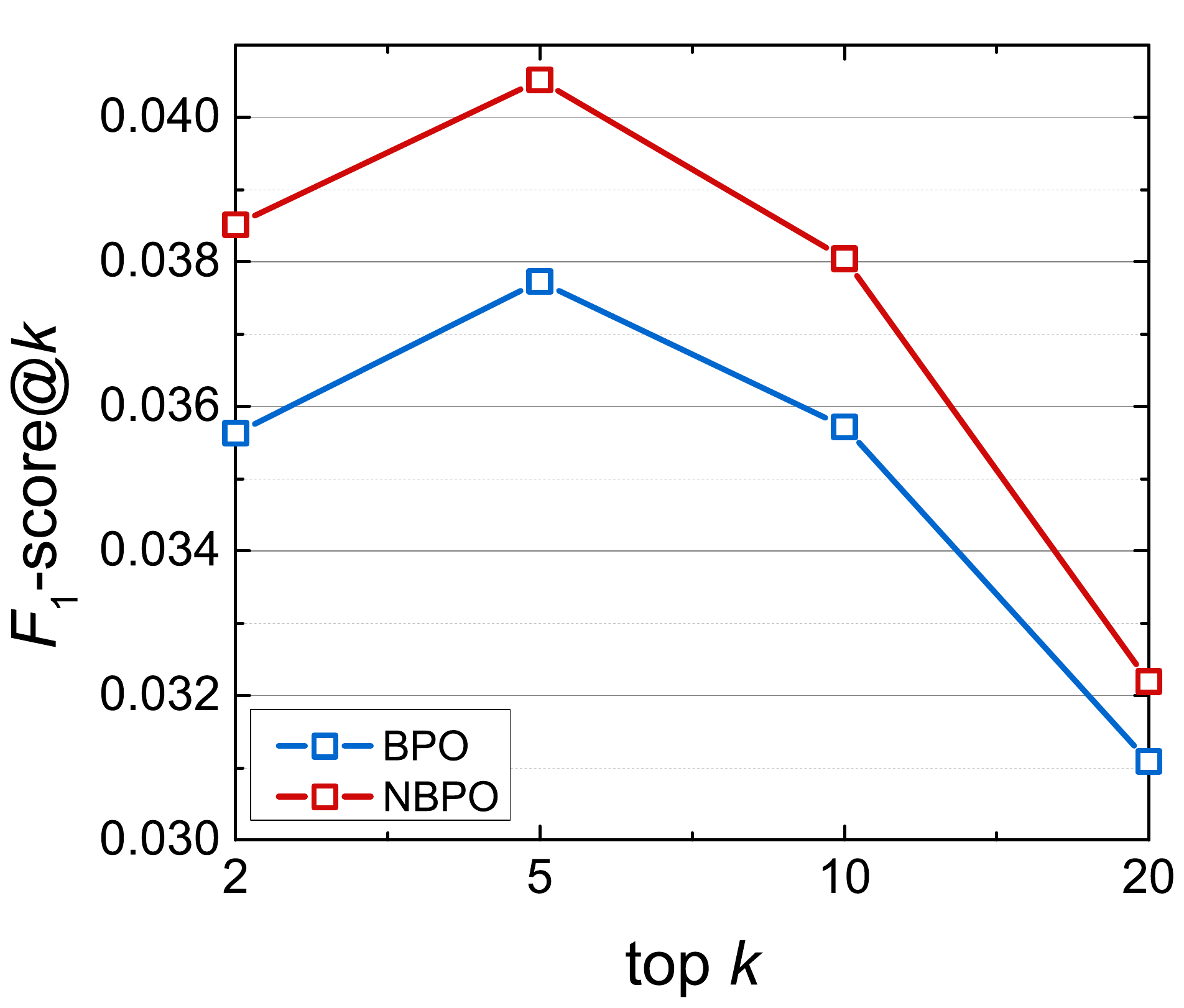}
        \label{subfig:amazon_RQ2_f1}
    }
    \hspace{-2mm}
    \subfigure[\textit{Movielens} --- $F_1$score@$k$]{
        \includegraphics[scale = 0.2]{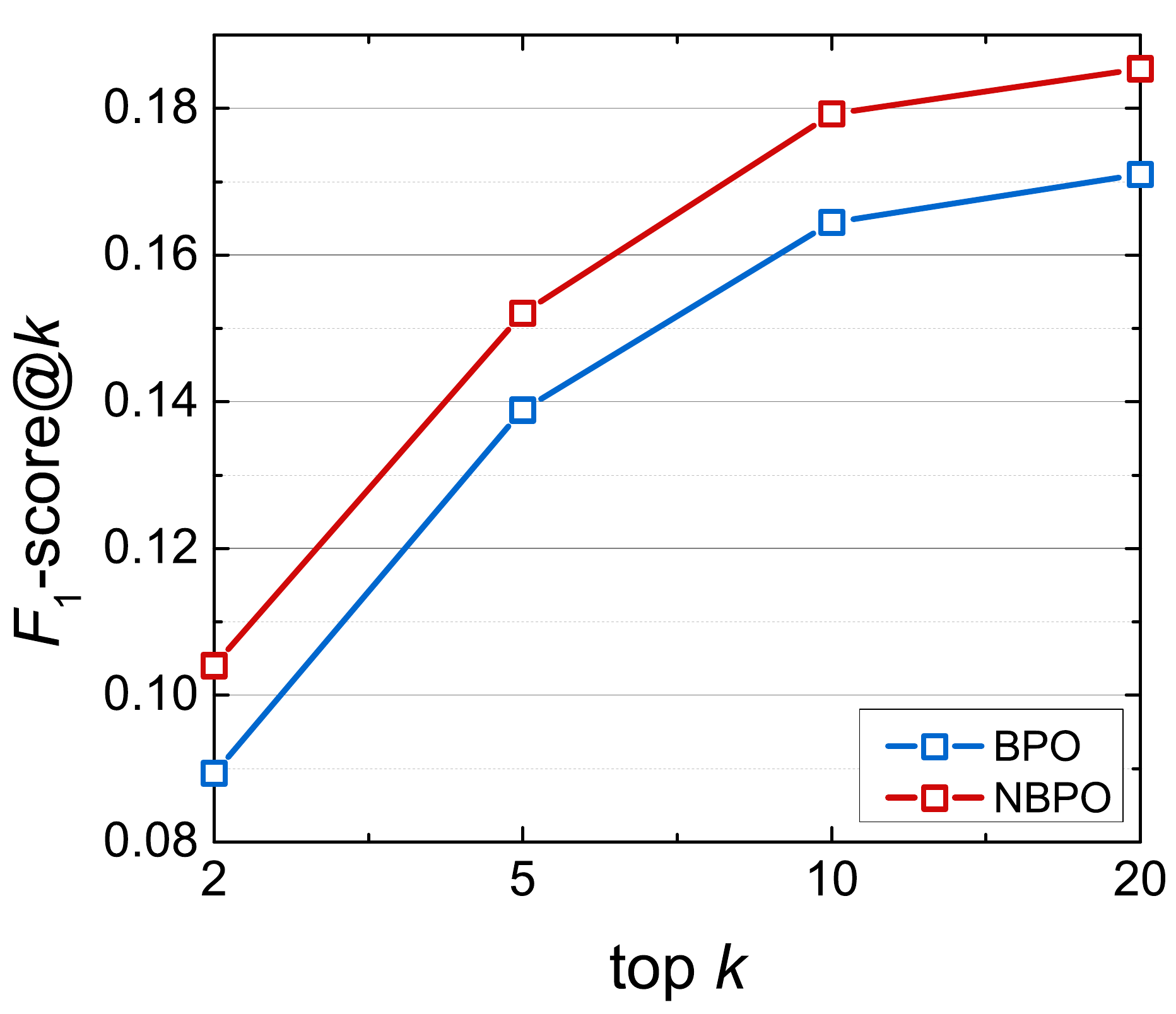}
        \label{subfig:movielens_RQ2_f1}
    }
    \caption{Recommendation performances for RQ2 (test set).\vspace{-0mm}}
    \label{fig:RQ2}
\end{figure}

In this subsection, we validate the effectiveness of exploring noisy-label robust learning in recommendation tasks. NBPO is compared against the basic optimization method BPO and the performances are reported in Figure \ref{fig:RQ2}. By modifying noisy labels, NBPO gains considerable accuracy improvement. NBPO outperforms BPO 8.08\% and 16.49\% for the best case on \textit{Amazon} and \textit{Movielens} datasets.

\begin{figure}[ht]
    \setlength{\abovecaptionskip}{2mm}
    \centering
    \subfigure[\textit{Amazon} --- $F_1$score@2]{
        \includegraphics[scale = 0.255]{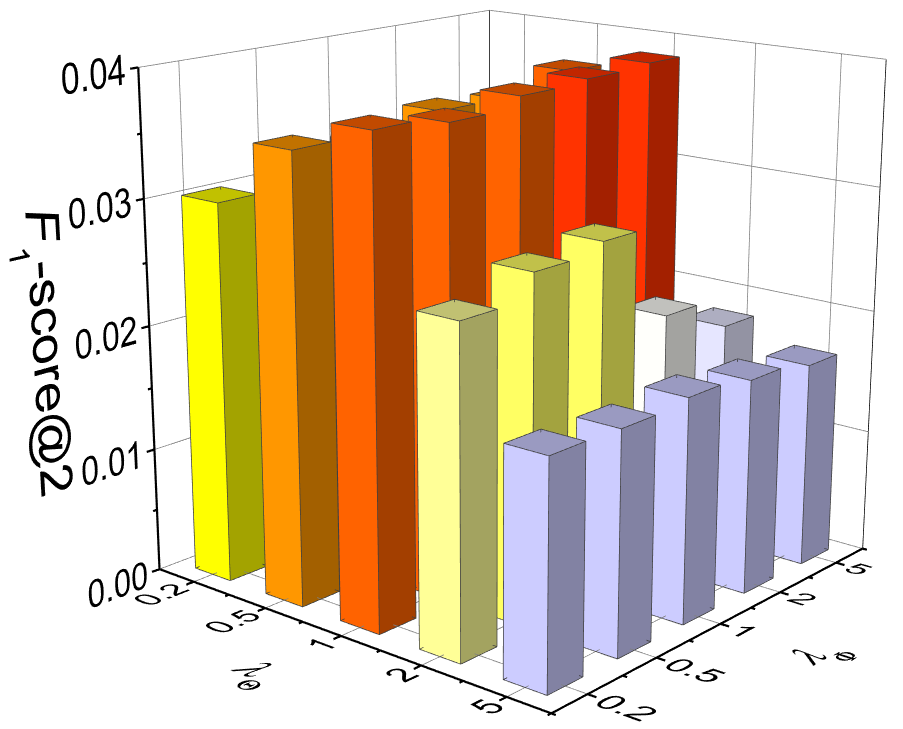}
        \label{subfig:amazon_lr_lf_f1}
    }
    \hspace{-2mm}
    \subfigure[\textit{Movielens} --- $F_1$score@2]{
        \includegraphics[scale = 0.255]{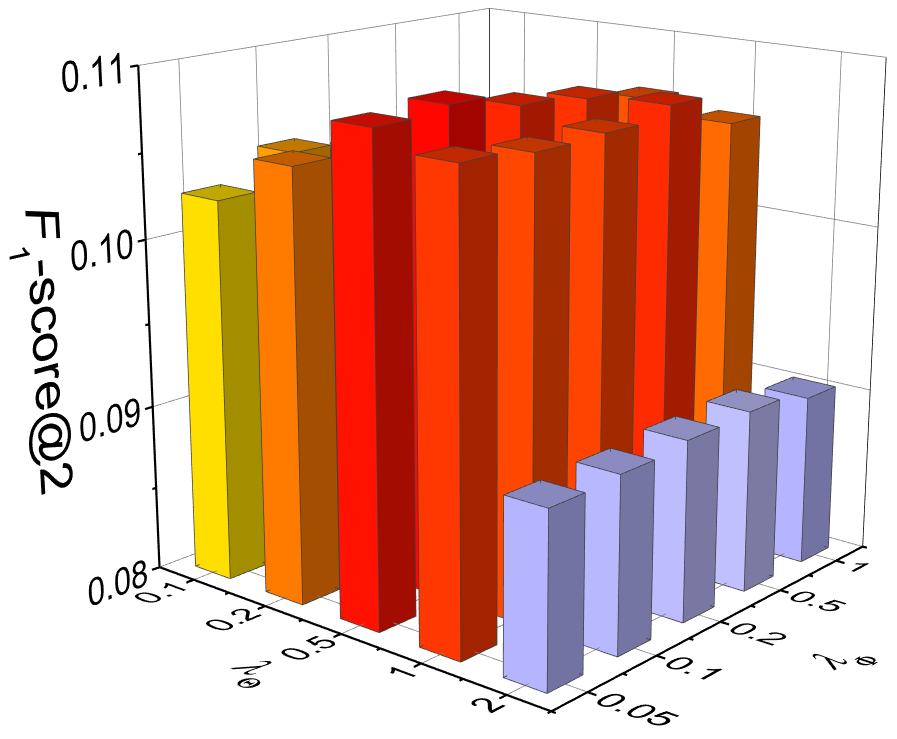}
        \label{subfig:movielens_lr_lf_f1}
    }
    \caption{Impact of regulation coefficients $\lambda_{\bm{{\rm \Theta}}}$ and $\lambda_{\bm{{\rm \Phi}}}$ (validation set).\vspace{-1mm}}
    \label{fig:lr_lf}
\end{figure}

In this subsection, we also show some details of NBPO tuning. We analyze the sensitivity of NBPO with varying regulation coefficients $\lambda_{\bm{{\rm \Theta}}}$ and $\lambda_{\bm{{\rm \Phi}}}$ in Figure \ref{fig:lr_lf}. As illustrated, when $\lambda_{\bm{{\rm \Theta}}}=1$ and $\lambda_{\bm{{\rm \Phi}}}=5$ on \textit{Amazon} and $\lambda_{\bm{{\rm \Theta}}}=0.5$ and $\lambda_{\bm{{\rm \Phi}}}=0.1$ on \textit{Movielens} dataset, NBPO achieves the best performance. From Figure \ref{fig:lr_lf} we can observe that NBPO is more sensitive with $\lambda_{\bm{{\rm \Theta}}}$ than with $\lambda_{\bm{{\rm \Phi}}}$, that is possibly because NBPO models users' preference with parameters ${\bm{{\rm \Theta}}}$ while optimizes with ${\bm{{\rm \Phi}}}$, thus ${\bm{{\rm \Theta}}}$ contribute to the performance more directly.

\begin{figure}[ht]
    \setlength{\abovecaptionskip}{2mm}
    \centering
    \subfigure[\textit{Amazon} --- $F_1$score@2]{
        \includegraphics[scale = 1]{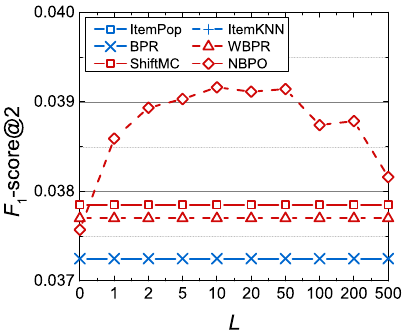}
        \label{subfig:amazon_L_f1}
    }
    \hspace{-2.5mm}
    \subfigure[\textit{Movielens} --- $F_1$score@2]{
        \includegraphics[scale = 1]{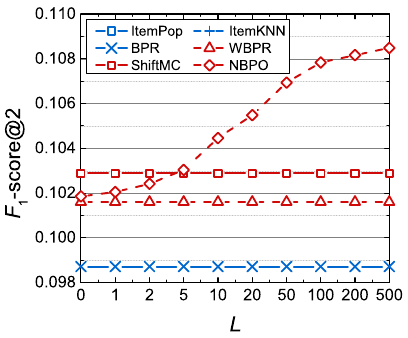}
        \label{subfig:Movielens_L_f1}
    }
    \caption{Impact of latent dimension number $L$ (validation set).}
    \label{fig:L}
\end{figure}

The impact of latent dimensions $L$ is illustrated in Figure \ref{fig:L}. When $L=10$ and 500, NBPO performs the best on \textit{Amazon} and \textit{Movielens} datasets, respectively.

A very interesting thing is that when $L=0$, our model also outperforms BPR. In this situation, Equation (\ref{equ:NBPO_gradient1}) degenerates to:
\begin{small}
\begin{eqnarray}
\label{equ:NBPO_gradient3}
\left.\begin{aligned}
\nabla_{\bm{{\rm \Theta}}} \mathcal{L}_s({\bm{{\rm \Phi}}},\!{\bm{{\rm \Theta}}})\!=\!\!\!&\sum_{(u,i)\in \mathcal{R}} \frac{1}{2} \sigma(-\hat{\bm{{\rm R}}}_{ui}({\bm{{\rm \Theta}}}))\frac{\partial {\bm{{\rm R}}}_{ui}({\bm{{\rm \Theta}}})}{\partial {\bm{{\rm \Theta}}}}\\
+\!\!\!&\sum_{(u,i)\notin \mathcal{R}} \frac{1}{2}\Big[1-3\sigma(\hat{\bm{{\rm R}}}_{ui}({\bm{{\rm \Theta}}}))\Big]\!\frac{\partial {\bm{{\rm R}}}_{ui}({\bm{{\rm \Theta}}})}{\partial {\bm{{\rm \Theta}}}} \!-\! \lambda_{\bm{{\rm \Theta}}} {\bm{{\rm \Theta}}}.
\end{aligned}
\right.
\end{eqnarray}
\end{small}In BPO and BPR, we train $\hat{\bm{{\rm R}}}_{ui}$ to $+\infty$ and train $\hat{\bm{{\rm R}}}_{uj}$ to $-\infty$ (where $i$ is the positive sample and $j$ is the negative sample). Updated by the gradient in Equation (\ref{equ:NBPO_gradient3}), $\hat{\bm{{\rm R}}}_{ui}$ is trained to $+\infty$ while $\hat{\bm{{\rm R}}}_{uj}$ is trained to $\sigma^{-1}(1/3)$, where $\sigma^{-1}(\;)$ is the inverse function of $\sigma(\;)$. It is a simple way to weaken the negative samples. By setting $L=0$ in NBPO, we can also gain performance enhancement without any additional time and space consumption.

\subsection{Effectiveness of the Surrogate Likelihood Function and Surrogate Gradient (RQ3)}
In this subsection, we validate the effectiveness of our key optimization techniques --- the surrogate objective function and surrogate gradient-based SGD method, by comparing these three models:
\begin{itemize}
\item{\textbf{NBPO-o:} We optimize the original log likelihood function ($\mathcal{L}({\bm{{\rm \Phi}}},{\bm{{\rm \Theta}}})$ shown in Equation (\ref{equ:NBPO_loglikelihood})) with conventional SGD in this NBPO-o model.}

\item{\textbf{NBPO-s:} We optimize the surrogate likelihood function ($\mathcal{L}_s({\bm{{\rm \Phi}}},{\bm{{\rm \Theta}}})$ shown in Equation (\ref{equ:NBPO_surrogate_log_likelihood})) with conventional SGD in this NBPO-s model.}

\item{\textbf{NBPO-ss:} We optimize the surrogate likelihood function ($\mathcal{L}_s({\bm{{\rm \Phi}}},{\bm{{\rm \Theta}}})$ shown in Equation (\ref{equ:NBPO_surrogate_log_likelihood})) with the surrogate gradient introduced in Equation (\ref{equ:NBPO_gradient1}) in this NBPO-ss model, which is our final NBPO model.}
\end{itemize}

\begin{figure}[ht]
    \setlength{\abovecaptionskip}{2mm}
    \centering
    \subfigure[\textit{Amazon}, $F_1$-score@$k$]{
        \includegraphics[scale = 0.21]{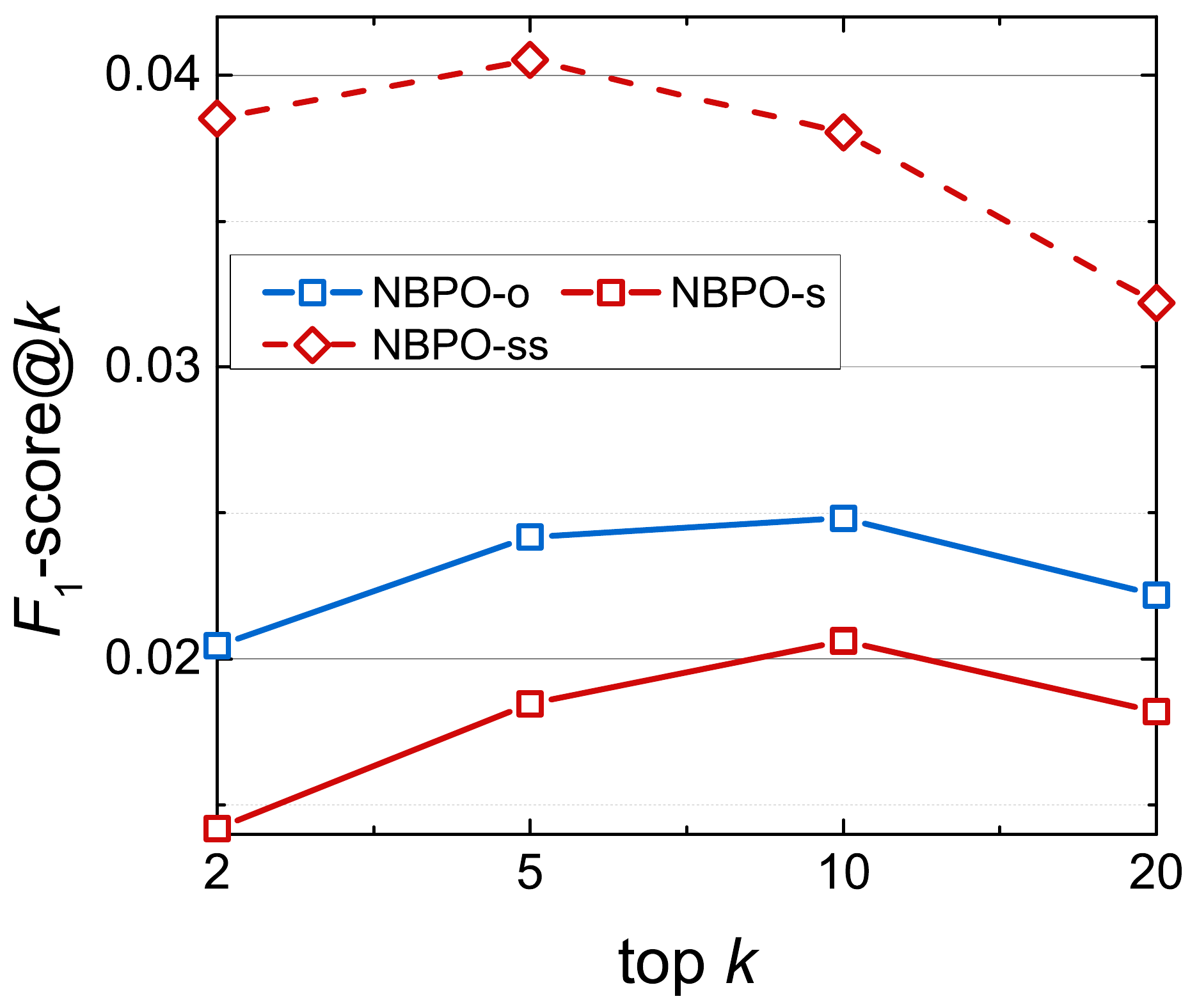}
        \label{subfig:amazon_RQ3_f1}
    }
    \hspace{-2mm}
    \subfigure[\textit{Movielens}, $F_1$-score@$k$]{
        \includegraphics[scale = 0.21]{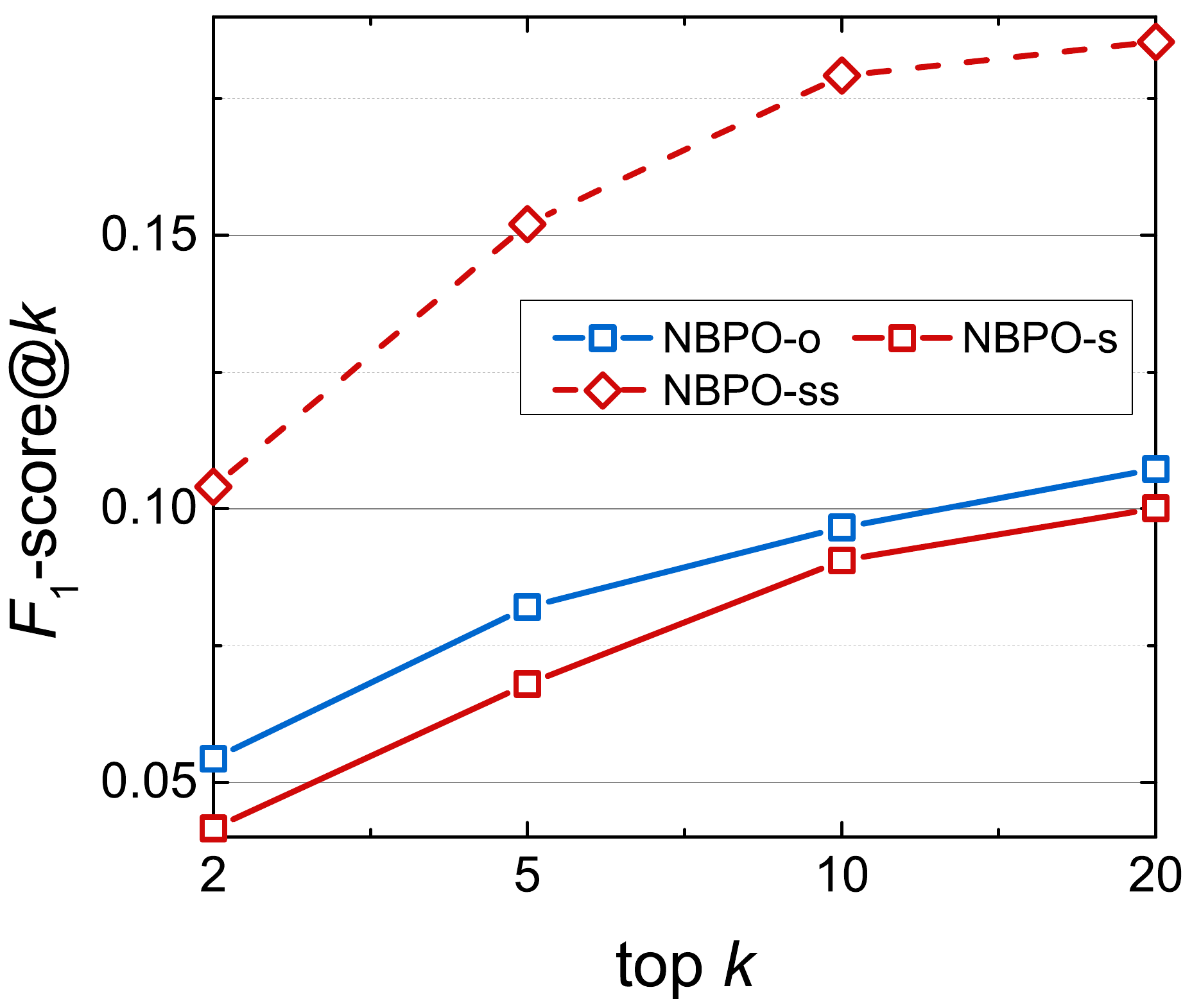}
        \label{subfig:movielens_RQ3_f1}
    }
    \caption{Recommendation performances for RQ3 (test set).}
    \label{fig:RQ3}
\end{figure}

Figure \ref{fig:RQ3} shows our NBPO models optimized by different methods. As we can see, NBPO-o performs pretty bad since it cannot be optimized well with vanilla MSGD. NBPO-s suffers from the ``gradient vanishing'' problem and performs the worst. To deal with the new issue, we further propose the surrogate gradient to update the model and optimization parameters. Enhanced by the surrogate likelihood function and the surrogate gradient, NBPO-ss performs the best in these three models on all metrics and all datasets.

\section{Conclusion and Future Work}
\label{sec:conclusion}
In this paper, we investigated the effectiveness of the noisy-label robust learning in recommendation domain. We first proposed BPO as our basic optimization method which maximizes the likelihood of observations, and then devised the NBPO model by exploring the noisy-label robust learning in BPO. In NBPO, we constructed the maximum likelihood estimator with the likelihood of users' preference and the likelihood of label flipping, and then estimated model parameters (user and item embeddings) and optimization parameters (label flipping likelihoods) by maximizing the estimator. To deal with the oversize issue of optimization parameters, we represented the likelihood matrix with MF. 

To be extensible to deep models for real-world applications and to improve the efficiency, we proposed a SGD-based optimization method. However vanilla SGD shows unsatisfactory performance in optimizing our maximum likelihood estimator. To address this gap, we maximized the lower bound of the original objective function inspired by EM algorithm, and we designed surrogate function and surrogate gradient for updating. Extensive experiments on challenging real-world datasets show that our model can improve the sampling quality and outperforms state-of-the-art models significantly.

For future work, we have interests in validating the effectiveness of NBPO with some advanced models, such as Factorization Machine (FM) \cite{FM} and some deep structures like Neural Matrix Factorization (NMF) \cite{NCF} and Attentive Collaborative Filtering (ACF) \cite{he_Attentive}. To handle the incompatibility issue of BPR, we proposed BPO as the basic optimization method, yet it shows weaker performance compared against BPR. We will devise a stronger basic optimization method, or combine noisy-label robust learning with other widely used optimization methods to improve the performance. Finally, noting that PU data is common in information retrieval tasks, such as text retrieval, web search, social network, etc., we have interests in extending the proposed optimization strategy to these fields.

\newpage

\newpage
\bibliographystyle{ACM-Reference-Format}


\end{document}